%% file: main.tex
\documentclass{ieeeaccess}
\usepackage{cite}
\usepackage{amsmath,amssymb,amsfonts}
\usepackage{algorithmic}
\usepackage{graphicx}
\usepackage{textcomp}
\usepackage[caption=false,font=footnotesize]{subfig}
\DeclareMathOperator{\Tr}{Tr}

\usepackage{etoolbox}
\makeatletter
\providecommand{\xfigwd}{\the\linewidth}

\patchcmd{\@makecaption}
  {\ifdim \xfigwd >\columnwidth}
  {\ifdim \xfigwd >\textwidth}
  {}{}
\makeatother
\makeatletter
\let\orig@fontseries\fontseries
\renewcommand{\fontseries}[1]{%
  \ifx#1n%
    \orig@fontseries{m}%
  \else
    \orig@fontseries{#1}%
  \fi}
\makeatother

\usepackage{array}
\usepackage{tabularx}

\newcolumntype{C}{>{$}c<{$}}

\usepackage[colorlinks,urlcolor=blue,linkcolor=blue,citecolor=blue]{hyperref}
\expandafter\def\expandafter\UrlBreaks\expandafter{\UrlBreaks\do\/\do\*\do\-\do\~\do\'\do\"\do\-}
\usepackage{color}
\usepackage{booktabs}
\usepackage{makecell}
\usepackage{cleveref}
\crefname{equation}{Eq.}{Eqs.}

\newcommand{\cc}{\mathrm{cc}}
\newcommand{\ent}{\mathrm{ent}}
\newcommand{\srv}{\mathrm{srv}}

\newcommand{\Tcc}{T_{\cc}}
\newcommand{\Tent}{T_{\ent}}
\newcommand{\Tsrv}{T_{\srv}}

\newcommand{\ncc}{n_{\cc}}

\def\BibTeX{{\rm B\kern-.05em{\sc i\kern-.025em b}\kern-.08em
    T\kern-.1667em\lower.7ex\hbox{E}\kern-.125emX}}
\begin{document}
\history{Date of publication xxxx 00, 0000, date of current version xxxx 00, 0000.}
\doi{10.1109/TQE.2020.DOI}

\title{Telemetry-Based Server Selection in the Quantum Internet via Cross-Layer Runtime Estimation}
\author{
    \uppercase{Masaki Nagai}\authorrefmark{1}, 
    \uppercase{Hideaki Kawaguchi}\authorrefmark{1},
    \uppercase{Shin Nishio}\authorrefmark{1,2},
    and
    \uppercase{Takahiko Satoh}\authorrefmark{3}
}
\address[1]{Graduate School of Science and Technology, Keio University, Yokohama, Kanagawa 223-8522, Japan (email: masaki0818nagai@keio.jp, hikawaguchi@keio.jp, shin.nishio@keio.jp)}
\address[2]{Department of Physics \& Astronomy, University College London, London, WC1E 6BT, United Kingdom}
\address[3]{Faculty of Science and Technology, Keio University, Yokohama, Kanagawa 223-8522, Japan (email: satoh@ics.keio.ac.jp)}
\tfootnote{
This work was supported by JST Moonshot R\&D Grant Number JPMJMS226C.
SN was also supported by the New Energy and Industrial Technology Development Organization (NEDO) JPNP23003 and the JSPS Overseas Research Fellowship.
TS was also supported by MEXT KAKENHI Grant Number 22K1978.
}
\markboth
{Nagai \headeretal: Telemetry Based Server Selection in the Quantum Internet via Cross-Layer Runtime Estimation}
{Nagai \headeretal: Telemetry Based Server Selection in the Quantum Internet via  Cross-Layer Runtime Estimation}
\corresp{Corresponding author: Masaki Nagai (email: masaki0818nagai@keio.jp).}

\begin{abstract}
The Quantum Internet will allow clients to delegate quantum workloads to remote servers over heterogeneous networks, but choosing the server that minimizes end-to-end execution time is difficult because server processing, feedforward classical communication, and entanglement distribution can overlap in protocol-dependent ways and shift the runtime bottleneck.
We propose $T_{\max}$, a lightweight runtime score that sums
coarse telemetry from multiple layers to obtain a conservative ranking for online server selection without calibrating weights for each deployment.
Using NetSquid discrete-event simulations of a modified parameter-blind VQE (PB-VQE) workload, we evaluate $T_{\max}$ on pools of 10{,}000 heterogeneous candidates (selecting among up to 100 per decision) across crossover and bottleneck-dominated regimes, including temporal jitter scenarios and jobs with multiple shots.
$T_{\max}$ achieves single-digit mean regret normalized by the oracle (below 10\%) in both regimes and remains in the single-digit range under classical communication latency jitter for multi-shot jobs, while performance degrades for single-shot jobs under severe jitter.
To connect performance to deployment planning, we derive an operating map based on requirements relating distance and entanglement rate requirements to protocol level counts, quantify how simple multiuser contention shifts the crossover, and use Sobol global sensitivity analysis to identify regime-dependent bottlenecks.
These findings suggest that simple cross-layer telemetry can enable practical server selection while providing actionable provisioning guidance for emerging Quantum Internet services.
\end{abstract}

\titlepgskip=-15pt

\maketitle
\input{texs/intro}

\input{texs/background}

\input{texs/problem}

\input{texs/eval_bvqe}
\input{texs/eval_bottleneck}
\input{texs/conclusion}
\input{texs/appendix}

\section*{Data and Code Availability}
All NetSquid simulation codes, configuration files, and analysis notebooks are publicly accessible at \url{https://github.com/masakivv/Server-Selection-in-the-Quantum-Internet.git}.

\bibliographystyle{IEEEtran}
\bibliography{IEEEabrv,main}

\EOD

\end{document}

%% file: texs/intro.tex
\section{Introduction} \label{sec:intro}

Quantum computers are expected to outperform classical devices on specific problems~\cite{shor1994algorithms,grover1996fast}, and the Quantum Internet is emerging as a key enabler for distributed computation and secure applications~\cite{RoadAhead, Kimble2008quantum, Simon2017global}. The foundations of quantum networking were laid in the 1990s with quantum teleportation, enabling state transfer via shared entanglement and classical communication~\cite{PhysRevLett.70.1895}, and with quantum repeaters, which introduced entanglement swapping and purification for long-distance communication~\cite{PhysRevLett.81.5932}.
These breakthroughs established the basis for modern entanglement distribution and repeater-based networks.

Recent experimental and theoretical advances have brought this vision closer to reality. Multi-node quantum networks have been realized over metropolitan fiber infrastructures~\cite{liu2024creation}, marking key milestones toward scalable implementations. In parallel, requirements for practical repeater deployment have been quantified both for processing-node architectures on real-world fiber grids~\cite{avis2023requirements} and for upgrading trusted-node links into long repeater chains~\cite{da2024requirements}. 

At the application layer, clients must select among heterogeneous remote quantum servers to minimize end-to-end runtime.
This problem is central to practical Quantum Internet applications.
In such applications, end-to-end runtime is determined not only by quantum processing but also by feedforward classical communication latency and entanglement distribution, each of which can lie on the critical path. 
Since whether these components overlap is protocol-dependent and bottlenecks can switch across regimes as distance, entanglement distribution rate, and server-side performance vary, application-layer server selection must rely on a lightweight telemetry-based ranking metric rather than detailed per-candidate modeling, active probing, or online learning.

Prior work has made substantial contributions from multiple directions. On the protocol side, blind quantum computation~(BQC) enables secure delegated computation with limited client capabilities~\cite{broadbent2009universalBFK,morimae2013blind}. 
In parallel, the variational quantum eigensolver (VQE) has emerged as the leading candidate for near-term applications, with demonstrations across platforms and extensive methodological studies~\cite{originVQE,peruzzo2014variational,VQEbestpractice}. 
Parameter-blind VQE (PB-VQE) has also been proposed to support secure and distributed use cases~\cite{BlindparametaVQE}. 
At the network and systems level, repeater architectures have been systematically compared to clarify distance-dependent performance trade-offs~\cite{muralidharan2016}, while utility-based frameworks connect network resources to socio-economic value~\cite{lee2024quantum}. 
In parallel, concrete protocol-stack decompositions and link-layer abstractions have been proposed and experimentally validated to expose entanglement delivery as an application-facing service, motivating lightweight cross-layer telemetry for runtime-aware decisions~\cite{Dahlberg2019linklayer,Pompili2022entanglementdelivery,KozlowskiDahlbergWehner2020designing,DelleDonne2025qnodeos}.

These efforts underscore the rapid progress of the field. Nevertheless, from the perspective of application-layer server selection, the literature still lacks a unified model that couples client, network, and server parameters to predict end-to-end runtime and guide server selection. In particular, three gaps persist: (i) the absence of a cross-layer runtime model capturing both computational and communication delays, (ii) the lack of a lightweight predictor that enables online ranking with minimal overhead, and (iii) an incomplete understanding of which physical and logical factors dominate end-to-end performance across operating regimes.

To address these gaps, we make three contributions. 
First, we formulate application-layer server selection as a cross-layer ranking problem under unknown overlap between server processing, feedforward classical communication, and entanglement distribution, and define two extreme-overlap runtime estimates.
Second, we propose a telemetry-only, tuning-free selection rule that ranks candidates by the conservative serialized estimate $T_{\max}$. This conservative choice avoids making overlap assumptions and provides a simple overlap-agnostic score for online ranking.
Third, using NetSquid discrete-event simulations~\cite{coopmans2021netsquid} of an implementation-oriented PB-VQE benchmark at scale ($10{,}000$-candidate pools and up to 100 candidates per decision), we quantify oracle-normalized regret across regimes (including scenario-based jitter) and connect performance to deployment planning via 
(i) a requirement-based $(d, R)$ operating map that captures protocol-level counts,
(ii) a simple multiuser contention scaling through $R_{\mathrm{eff}} = R/U$, and
(iii) a regime map of dominant bottlenecks based on Sobol total-effect sensitivity analysis~\cite{sobol1990sensitivity}.

We begin by reviewing the Quantum Internet setting and the delegated workload that motivates our focus on cross-layer delays (Sec.~\ref{sec:back}). 
We then formalize latency components and define two extreme-overlap runtime estimates that lead to a robust server selection rule (Sec.~\ref{sec:problem}). 
The proposed policy is evaluated via NetSquid discrete-event simulations across heterogeneous candidate sets and operating regimes, including robustness under temporal jitter (Sec.~\ref{sec:evaluation}). 
We complement these results with a requirement-based operating map and global sensitivity analysis to identify dominant bottlenecks and extract deployment-relevant priorities  (Sec.~\ref{sec:sobol}). 
We conclude with limitations and directions toward load-aware predictors for multiuser deployments (Sec.~\ref{sec:conclusion}).

%% file: texs/background.tex
\section{Background}\label{sec:back}
\subsection{Overview of the Quantum Internet}
\begin{figure}[htbp]
  \centering
  \includegraphics[width=1\linewidth]{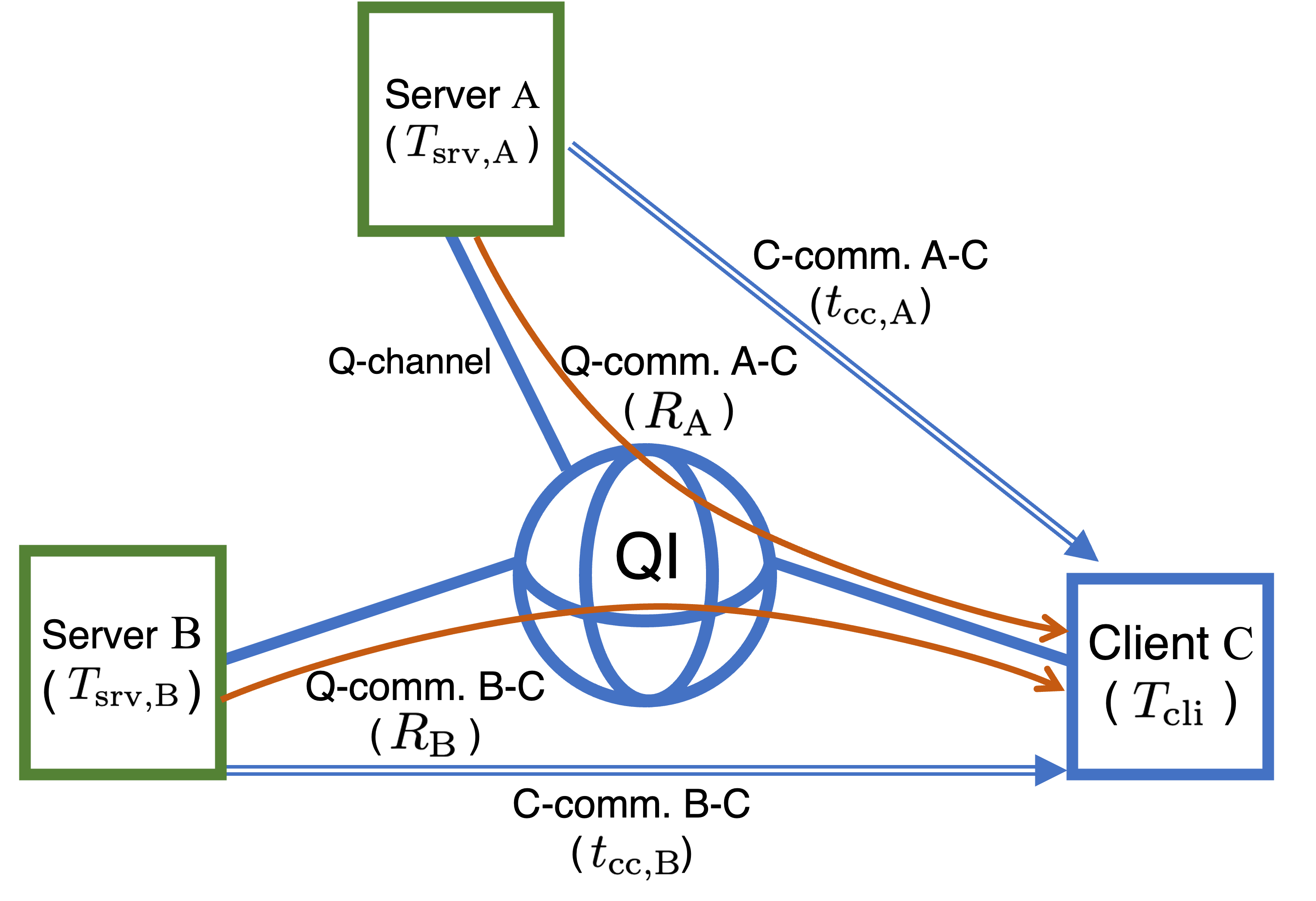} 
  \caption{Client–network–server schematic for the server selection problem, instantiated for the PB-VQE workflow. Two candidate servers (A and B) communicate with the client C over quantum channels through the shared Quantum Internet (QI). The quantum communication links (A–C and B–C) are annotated with the effective entanglement distribution rate $R$, while the classical feedforward communication latency $t_{\mathrm{cc}}$ depends on the physical channel length $d$ (propagation-limited model). Server nodes are parameterized by the server-side processing time $T_{\mathrm{srv}}$, and the client node by $T_{\mathrm{cli}}$. Parenthesized symbols indicate the telemetry components used in the runtime decomposition $(T_{\mathrm{cli}}, T_{\mathrm{srv}}, T_{\mathrm{cc}}, T_{\mathrm{ent}})$.}
  \label{fig:QI}
\end{figure}

The Quantum Internet, which distributes entanglement between remote nodes, has emerged as a foundational infrastructure for quantum cloud computing and secure quantum communication. Unlike the classical Internet, which only transmits classical bits, the Quantum Internet employs quantum channels for state transfer and entanglement distribution in addition to classical communication~\cite{RoadAhead}. A client with limited quantum capability can connect to multiple remote servers through the Quantum Internet, where both quantum and classical channels are indispensable for the application workflow (Fig.~\ref{fig:QI}).  

\subsection{Blind Quantum Computation Protocols}
Research on BQC has produced a range of protocols tailored to different client capabilities. Canonical examples include the BFK protocol ~\cite{broadbent2009universalBFK}, in which the client prepares single qubits with random rotations, and the MF protocol~\cite{morimae2013blind}, in which the client only performs measurements. 
These protocols ensure computational blindness, but can incur substantial overhead on the client--server critical path, making end-to-end runtime sensitive to both network latency and entanglement distribution.
Complementary approaches explore alternative trust assumptions for secure delegation, such as quantum trusted execution environments, which can reduce interaction overhead at the cost of different threat models~\cite{Ma2022qenclave}.

\subsection{Parameter-Blind Variational Quantum Eigensolver}
The variational quantum eigensolver is a leading near-term algorithm, but in delegated use, its classical--quantum loop demands many shots. The parameter-blind VQE (PB-VQE) addresses privacy and communication overhead by fixing the server circuit while the client updates hidden parameters~\cite{BlindparametaVQE}. We further enhance the PB-VQE protocol by integrating quantum teleportation, which increases robustness against channel loss. 

We adopt PB-VQE as an implementation-oriented workload for studying client--server execution over the Quantum Internet. Each shot combines a fixed server-side circuit with a small set of feedforward updates at the client, which makes the end-to-end runtime sensitive to both hardware and network conditions. The fixed circuit allows us to parameterize server-side computational performance mainly by gate duration and readout, while the network contribution is determined by classical communication latency over fiber and the entanglement distribution rate required to supply Bell pairs.

We use this workload because it exposes the cross-layer delays that drive end-to-end runtime while keeping protocol-level counts on the critical path fixed, and we provide protocol details in Appendix~\ref{sec:pbvqe}.

%% file: texs/problem.tex
\section{Problem Formulation} \label{sec:problem}
\subsection{Workflow and latency components}
\begin{figure*}[t]
    \centering
    \includegraphics[width=1\linewidth]{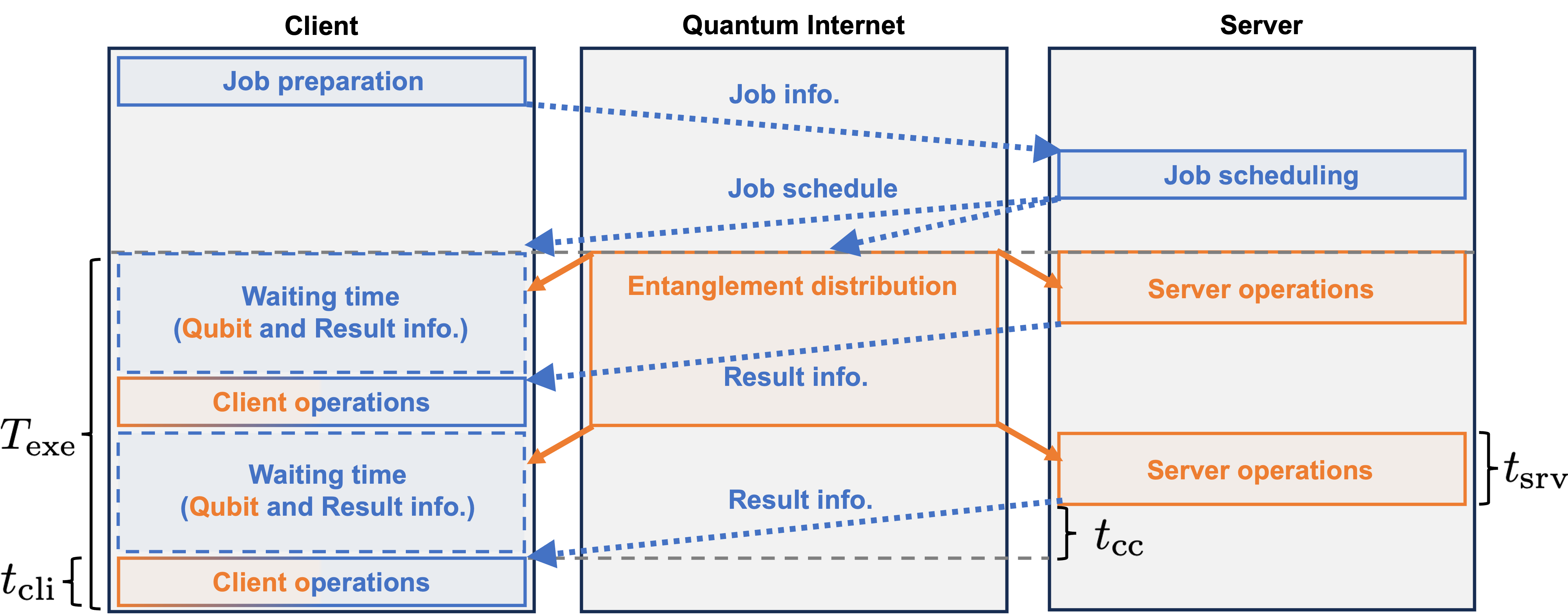}
    \caption{Protocol diagram of a client--network--server workflow over the Quantum Internet. Blue denotes classical operations/communication, and orange denotes quantum ones. Latency parameters $T_{\mathrm{srv}}, T_{\mathrm{cli}}, T_{\mathrm{cc}}, T_{\mathrm{ent}}$ denote server processing, client processing, classical communication, and entanglement distribution, respectively.}
    \label{fig:protocol-diagram}
\end{figure*}

The end-to-end workflow of a Quantum Internet application is a network-mediated client--server interaction.
The client prepares and submits a job, the network distributes entanglement, and the server executes quantum operations while exchanging classical messages with the client (Fig.~\ref{fig:protocol-diagram}).
The end-to-end execution time $T_{\mathrm{exe}}$ reflects the combined delays of multiple layers, where the dominant bottleneck shifts across regimes. This regime-dependent behavior necessitates a cross-layer runtime formulation.

\label{sec:runtime}
The end-to-end runtime can be decomposed into four latency components: server processing ($T_{\mathrm{srv}}$), client processing ($T_{\mathrm{cli}}$), classical communication ($T_{\mathrm{cc}}$), and entanglement distribution ($T_{\mathrm{ent}}$).
We formalize each component as follows:
\begin{align}
T_{\mathrm{srv}} &= \sum t_{\mathrm{srv}}, \label{eq:Tsrv}\\
T_{\mathrm{cc}} &= n_{\mathrm{cc}}\, t_{\mathrm{cc}}, \label{eq:Tcc}\\
T_{\mathrm{ent}} &= \frac{K}{R}, \label{eq:Tent}\\
T_{\mathrm{cli}} &= \sum t_{\mathrm{cli}}. \label{eq:Tcli}
\end{align}
Here, $T_{\mathrm{srv}}$ and $T_{\mathrm{cli}}$ denote the critical-path execution time of the fixed server circuit under the assumed scheduling model (which can be exposed
as coarse-grained telemetry or derived from device specifications for a fixed benchmark circuit). $n_{\mathrm{cc}}$ is the number of one-way classical messages on the critical path,
$t_{\mathrm{cc}}$ is the one-way latency per message, and $K$ is the number of Bell pairs consumed.
$T_{\mathrm{cc}}=n_{\mathrm{cc}}\, t_{\mathrm{cc}}$ captures the serialized feedforward delay along the critical path, and
$T_{\mathrm{ent}}=K/R$ uses the effective end-to-end entanglement distribution rate $R$ at the target fidelity.
Appendix~\ref{sec:pbvqe} provides the protocol-specific mapping for PB-VQE, including how $(n_{\mathrm{cc}},K)$ arise and how
$T_{\mathrm{srv}}$ is obtained for the benchmark circuit.

In what follows, we model only the execution-phase latency that differentiates candidate servers and omit one-time job setup and scheduling overheads. These overheads are implementation-dependent and are treated as constant across candidates in our evaluation. Accordingly, we base the optimization on the end-to-end execution time $T_{\mathrm{exe}}$.

\subsection{Runtime Predictor}
\label{sec:runtime-model}

The end-to-end execution time $T_{\mathrm{exe}}$ of a Quantum Internet application is determined
by multiple latency components across the client--server stack. To obtain lightweight runtime
estimates suitable for online server ranking, we consider two limiting overlap regimes and define:
\begin{align}
T_{\max} &= T_{\mathrm{cli}} + T_{\mathrm{srv}} + T_{\mathrm{cc}} + T_{\mathrm{ent}}, \label{eq:tmax}\\
T_{\min} &= \max\{T_{\mathrm{cli}}, T_{\mathrm{srv}}, T_{\mathrm{cc}}, T_{\mathrm{ent}}\}. \label{eq:tmin}
\end{align}
Here, $T_{\max}$ corresponds to a fully serialized execution, whereas $T_{\min}$ corresponds to a perfectly
overlapped, bottleneck-limited execution. They represent opposite extremes of the overlap assumption.
We treat the $T_{\max}$-based policy as the proposed server selection method, while we use
$T_{\min}$ only as a reference baseline.

In practice, the realized overlap among these components is implementation- and load-dependent and is difficult to infer from coarse telemetry.
We therefore treat $T_{\max}$ as a conservative, monotonic score for online ranking, and validate its selection quality empirically in Sec.~\ref{sec:evaluation}.

As formalized in Sec.~\ref{sec:runtime}, we decompose the workflow into client processing ($T_{\mathrm{cli}}$),
server processing ($T_{\mathrm{srv}}$), classical communication latency ($T_{\mathrm{cc}}$), and entanglement distribution
($T_{\mathrm{ent}}$). These components do not compose additively in general. For example, feedforward dependencies
can serialize a subset of $T_{\mathrm{cc}}$ on the critical path, whereas $T_{\mathrm{ent}}$ may proceed concurrently with
local operations and even with parts of $T_{\mathrm{cc}}$ when dependencies and resources permit. As channel length,
entanglement distribution rate, and server-side computational performance differ across candidates, the dominant
latency component can shift across operating regimes, making server ranking based on a single component inconsistent.

\subsection{Server Selection Problem}
We restrict attention to the subset of candidate servers that satisfy the application-level accuracy requirement. 
Let $\mathcal{F}{\mathrm{acc}}$ denote the feasible candidate set after this offline filtering.
Given a scoring function $\mathrm{score}_i$, the client selects 
\begin{equation} 
i^* \in \arg\min_{i\in \mathcal{F}_{\mathrm{acc}}} \mathrm{score}_i, 
\end{equation} 
where $\mathcal{F}_{\mathrm{acc}}$ denotes the subset of candidates that satisfy the application-level accuracy requirement.
In this work, we assume that feasibility is evaluated offline and leave online accuracy estimation to future work.

%% file: texs/eval_bvqe.tex
\section{Performance Evaluation}\label{sec:evaluation}
\subsection{Experimental Setup}\label{sec:setup}
Building on the runtime predictors in Sec.~\ref{sec:problem}, we evaluate telemetry-based server selection policies via NetSquid discrete-event simulations across heterogeneous channel length, entanglement distribution rate, and server-side computational performance. 
NetSquid is one of several simulation frameworks for Quantum Internet software and performance studies. Related tools include SimulaQron, SeQUeNCe, and QuNetSim~\cite{DahlbergWehner2018simulaqron, Wu2021sequence, DiAdamo2021qunetsim}.
We consider a setting where a client selects among $M$ candidate servers with heterogeneous network conditions and server-side operation speeds. The client ranks candidates using the latency components {$T_{\mathrm{srv}}$, $T_{\mathrm{cc}}$, and $T_{\mathrm{ent}}$} under a fixed client implementation.

We parameterize each candidate server by the three scalars $(\kappa,d,R)$, which map to the telemetry components $(T_{\mathrm{srv}},T_{\mathrm{cc}},T_{\mathrm{ent}})$ via Eqs.~\eqref{eq:Tsrv}--\eqref{eq:Tent}.
Here, $\kappa$ scales all server-side gate and measurement durations (Table~\ref{tab:even}), $d$ is the channel length that determines classical communication latency, and $R$ is the end-to-end entanglement distribution rate.

The evaluation procedure for each regime is as follows:
\begin{enumerate}
\item Generate a pool of $10{,}000$ candidate servers by sampling $(d, R, \kappa)$ independently on a logarithmic scale. 
\item For each candidate-set size $M \in \{1, \ldots , 100\}$, draw $1{,}500$ instances by sampling $M$ servers from the pool. 
\item For each instance, compute telemetry-based scores {$T_{\mathrm{srv}},T_{\mathrm{cc}},T_{\mathrm{ent}}$} for every candidate, apply each selection policy, and simulate PB-VQE to obtain the realized runtime $T_{\mathrm{exe}}$ of the selected server.
\item For evaluation only, define the oracle-selected server as $\arg\min T_{\mathrm{exe}}$ within the instance and
report oracle-normalized regret
$r = (T_{\mathrm{exe}}^{(\text{policy})}-T_{\mathrm{exe}}^{(\text{oracle})})/T_{\mathrm{exe}}^{(\text{oracle})}$.
\end{enumerate}
We sample $(d, R, \kappa)$ independently to model a heterogeneous candidate pool in which the effective end-to-end entanglement service rate $R$ can be decoupled from physical distance $d$ by provisioning choices and differing network architectures (e.g., repeater chains and routing over distinct paths). Sec.~\ref{sec:operating-map} later reintroduces the physical coupling between $d$ and achievable entanglement supply through a requirement-based operating map (Fig.~\ref{fig:operating-map}).

For evaluation only, we obtain the oracle by simulating PB-VQE for all $M$ candidates in each instance to compute
$\arg\min T_{\mathrm{exe}}$. The oracle is not available to the online policy.
We benchmark the modified PB-VQE protocol on the H$_2$ molecular Hamiltonian~\cite{originVQE}, a standard test case for variational quantum algorithms.
Each shot requires $n_{\mathrm{cc}} = 5$ classical 
messages on the critical path and consumes $K = 5$ Bell pairs. 
These counts are protocol-specific and problem-dependent.

We model the one-way classical communication latency as
$t_{\mathrm{cc}} = d/v_{\mathrm{fiber}}$,
where $d$ is the channel length and
$v_{\mathrm{fiber}} \approx 2\times 10^5\,\mathrm{km/s}$
is the effective speed of light in optical fiber.
Substituting it into Eq.~\eqref{eq:Tcc} yields
$T_{\mathrm{cc}} = n_{\mathrm{cc}}\,t_{\mathrm{cc}} = n_{\mathrm{cc}}d/v_{\mathrm{fiber}}$.
We compute the entanglement distribution time from Eq.~\eqref{eq:Tent} as
$T_{\mathrm{ent}} = K/R$,
where $R$ is the end-to-end entanglement distribution rate.

We approximate the three latency components in~\cref{eq:Tsrv,eq:Tcc,eq:Tent} for a single shot of the PB-VQE benchmark as
\begin{equation}
\label{eq:even}
\bigl(T_{\mathrm{srv}},\,T_{\mathrm{cc}},\,T_{\mathrm{ent}}\bigr)
\approx
\left(9.48\,\kappa,\;0.025\,d,\;\frac{5000}{R}\right)\ \mathrm{ms}
\end{equation}
where gate durations are taken from recent trapped-ion experiments (IonQ Aria~\cite{IonqAriaperformance}) and $\kappa$ is an operation-duration scaling factor.
The prefactor $9.48\,\mathrm{ms}$ corresponds to the total duration of the fixed server-side circuit used in this benchmark (Appendix~\ref{sec:pbvqe}, Fig.~\ref{fig:pbvqe-circuit}). We sample the channel length $d$ in $\mathrm{km}$, the end-to-end entanglement distribution rate $R$ in $\mathrm{Hz}$, and the operation-duration scaling factor $\kappa$ independently by drawing each parameter uniformly on a logarithmic scale to generate a regime-specific pool of $10{,}000$ servers. In the Even regime, we sample $(d, R, \kappa)$ over the ranges listed in Table~\ref{tab:even}, which yields
$T_{\mathrm{srv}}, T_{\mathrm{cc}}, T_{\mathrm{ent}} \in [2.5, 25]~\mathrm{ms}$.
We choose the range of $\kappa$ so that $T_{\mathrm{srv}}$ spans the same order of magnitude as $T_{\mathrm{cc}}$ and $T_{\mathrm{ent}}$ in the Even regime, enabling balanced trade-offs among the three components.

\begin{table}[htbp]
\centering
\caption{Hardware parameters and variation ranges for the Even regime.}
\label{tab:even}
\small
\setlength{\tabcolsep}{4pt}
\renewcommand{\arraystretch}{1.05}
\begin{tabularx}{\columnwidth}{@{}
  >{\hsize=0.70\hsize\raggedright\arraybackslash}X
  >{\hsize=0.30\hsize\raggedright\arraybackslash}X
@{}}
\hline
\textbf{Parameter} & \textbf{Value} \\
\hline
1-qubit gate duration & $135\,\kappa\,\mu\mathrm{s}$ \\
2-qubit gate duration & $600\,\kappa\,\mu\mathrm{s}$ \\
Readout duration      & $200\,\kappa\,\mu\mathrm{s}$ \\
Operation-duration scaling factor $\kappa$ & $0.263$--$2.63$ \\
Entanglement distribution rate $R$ & $200$--$2000\,\mathrm{Hz}$ \\
Channel length $d$ & $100$--$1000\,\mathrm{km}$ \\
\hline
\end{tabularx}
\end{table}

Unless otherwise stated, we set the shot count to $S=10$.
Across the sweep ranges in Table~\ref{tab:even}, this choice places $T_{\mathrm{exe}}$ in the sub-second to few-second range.
In the deterministic model, $T_{\mathrm{srv}}$, $T_{\mathrm{cc}}$, and $T_{\mathrm{ent}}$ scale linearly with $S$, so the oracle ranking and the oracle-normalized regret change little with $S$.
We also report $S=1$ in Sec.~\ref{sec:result} to test sensitivity under temporal fluctuations.

\subsection{Policies and Overhead}
\label{sec:policy}
For online server selection, the client must make a selection with negligible decision overhead. We characterize a selection policy along three practical axes: (i) decision-time computation over $M$ candidates, (ii) additional learning/tuning, and (iii) observation budget, i.e., the telemetry required to score each candidate. 
We assume that $\{T_{\mathrm{srv}}, T_{\mathrm{cc}}, T_{\mathrm{ent}}\}$ are available as coarse-grained, periodically updated telemetry, rather than per-decision active measurements. Since the client and protocol are held fixed across candidates, $T_{\mathrm{cli}}$ is constant across candidates and is omitted from the decision variable.

We evaluate the proposed $T_{\max}$-based policy against the following baselines. Each policy assigns a telemetry-based runtime estimate $\mathrm{score}_i$ to candidate server $i$, and the client selects $i^*=\arg\min_i \mathrm{score}_i$. The scores are defined as follows:
\begin{align}
&\text{($T_{\max}$)}\quad \mathrm{score}_i =
T_{\mathrm{srv},i}+T_{\mathrm{cc},i}+T_{\mathrm{ent},i}, \\
&\text{($T_{\min}$)}\quad \mathrm{score}_i =
\max\{T_{\mathrm{srv},i},T_{\mathrm{cc},i},T_{\mathrm{ent},i}\}, \\
&\text{(Rank-sum)}\quad \mathrm{score}_i =
\sum_{k\in\{\mathrm{srv},\mathrm{cc},\mathrm{ent}\}} \mathrm{rank}(T_{k,i}), \\
&\text{(Weighted-sum)}\quad \mathrm{score}_i =
\sum_{k\in\{\mathrm{srv},\mathrm{cc},\mathrm{ent}\}} w_k T_{k,i}.
\end{align}

For Weighted-sum, we use Sobol-derived weights (computed offline as described in Sec.~\ref{sec:sobol}). 
The weights $w_k$ are precomputed from the Sobol sensitivity analysis in Sec.~\ref{sec:sobol} and then fixed across all instances.
We additionally include single-component indicators (server-side computational performance, classical communication latency, and entanglement distribution rate) and a random policy as simple reference baselines. Table~\ref{tab:policy_complexity} summarizes the per-decision complexity and additional learning requirements.

\begin{table}[htbp]
\centering
\caption{Decision-time complexity and additional learning for selection policies. $M$ is the number of candidate servers.
All policies assume access to $\{T_{\mathrm{srv}},T_{\mathrm{cc}},T_{\mathrm{ent}}\}$. Weighted-sum additionally requires offline calibration to obtain $w$.}
\label{tab:policy_complexity}
\setlength{\tabcolsep}{3pt}
\renewcommand{\arraystretch}{1.1}

\begin{tabularx}{\linewidth}{@{}
  >{\hsize=0.90\hsize\raggedright\arraybackslash}X
  >{\hsize=1.20\hsize\raggedright\arraybackslash}X
  >{\hsize=0.90\hsize\raggedright\arraybackslash}X
@{}}
\hline
\textbf{Policy} & \textbf{per-decision complexity} & \textbf{additional learning} \\
\hline
$T_{\min}$   & $\mathcal{O}(M)$, linear scan   & none \\
$T_{\max}$   & $\mathcal{O}(M)$, linear scan   & none \\
Rank-sum       & $\mathcal{O}(M\log M)$, sorting & none \\
Weighted-sum   & $\mathcal{O}(M)$, linear scan   & weight calibration \\
\hline
\end{tabularx}
\end{table}
\subsubsection*{Practical overhead beyond per-decision complexity}
Table~\ref{tab:policy_complexity} shows that both $T_{\max}$ and Weighted-sum can be evaluated in $\mathcal{O}(M)$ time at decision time. The practical difference is that Weighted-sum requires an offline, deployment-specific calibration step to determine the weight vector $w$ (e.g., from assumed operating priors or from labeled runs),
and the benefit of calibrated weights can diminish when the candidate population
 or network conditions shift (Sec.~\ref{sec:result} and Appendix~\ref{app:sobol}). In contrast,
 $T_{\max}$ uses coarse telemetry directly with no per-deployment weight calibration,
 making it a robust default when operating conditions are heterogeneous or time-varying.
\subsubsection*{Rationale for $T_{\max}$}
Because the degree of overlap among $T_{\mathrm{srv}}$, $T_{\mathrm{cc}}$, and $T_{\mathrm{ent}}$ depends on protocol details and runtime conditions (e.g., scheduling and jitter), it is generally not predictable from coarse-grained telemetry alone.
We therefore use the fully-serialized proxy $T_{\max}$ as a conservative, tuning-free ranking score: it is exact under serialized execution and remains monotone in each latency component, which makes it robust under bottleneck switching (Sec.~\ref{sec:bnswitch}).
In contrast, $T_{\min}$ corresponds to a perfectly-overlapped optimistic extreme and is included only as a reference baseline.

\subsection{Bottleneck switching}\label{sec:bnswitch}
\begin{figure*}[t]
  \centering
  \subfloat[Entanglement distribution rate ($R$)]{%
    \includegraphics[width=0.333\linewidth]{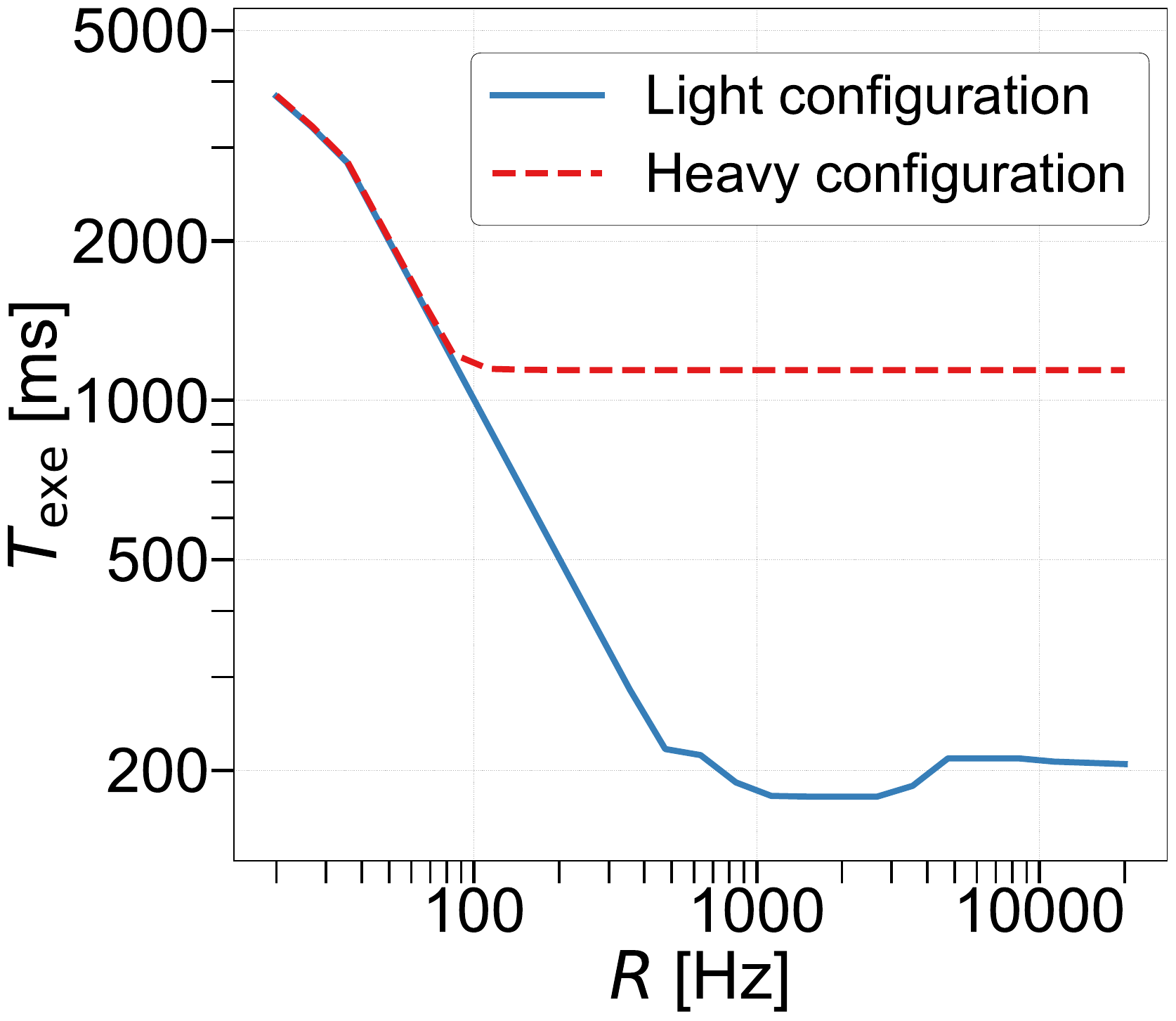}%
    \label{fig:two_anchor_sweep_R}%
  }\hfill
  \subfloat[Distance ($d$)]{%
    \includegraphics[width=0.333\linewidth]{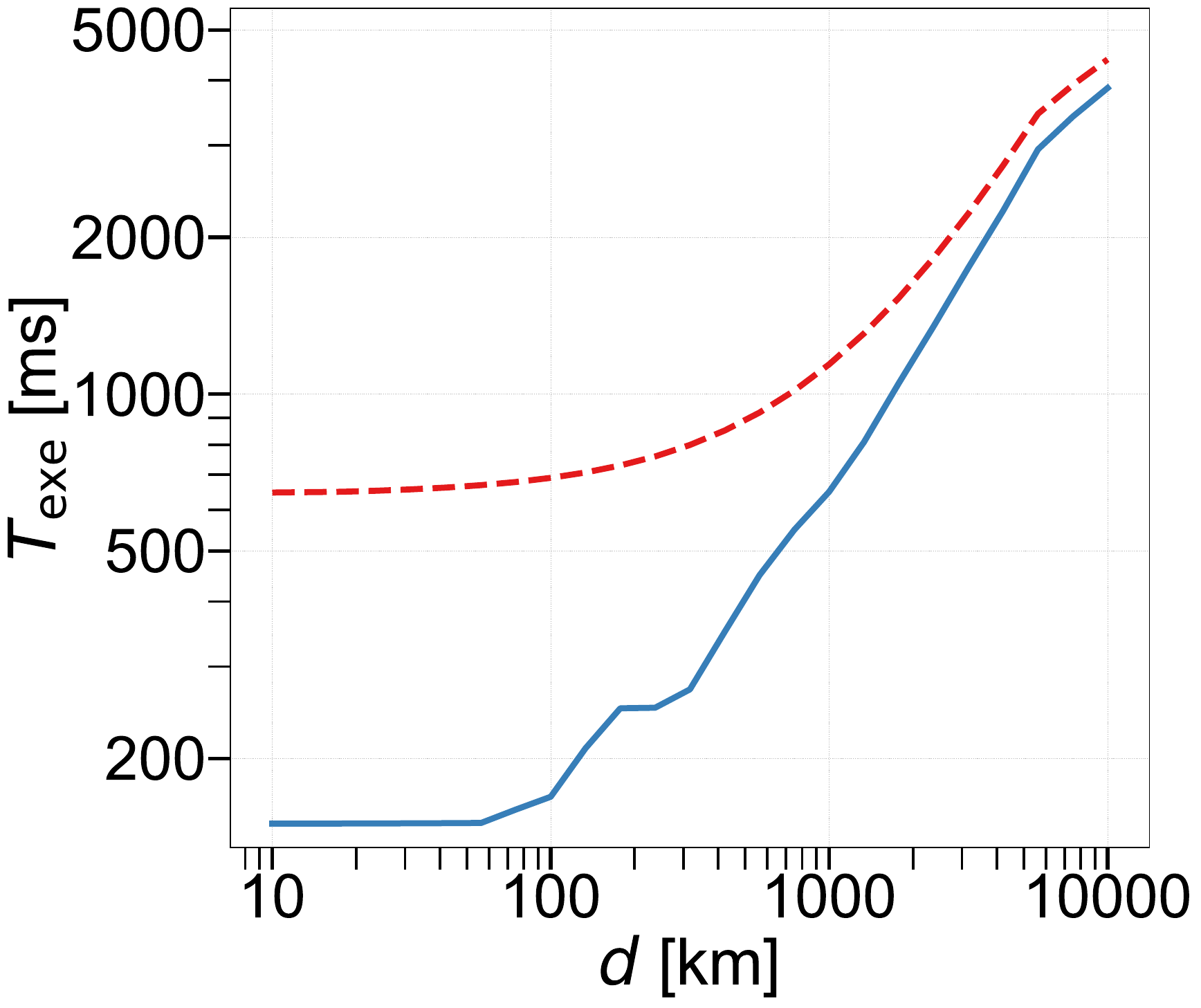}%
    \label{fig:two_anchor_sweep_d}%
  }\hfill
  \subfloat[Operation-duration scaling factor ($\kappa$)]{%
    \includegraphics[width=0.333\linewidth]{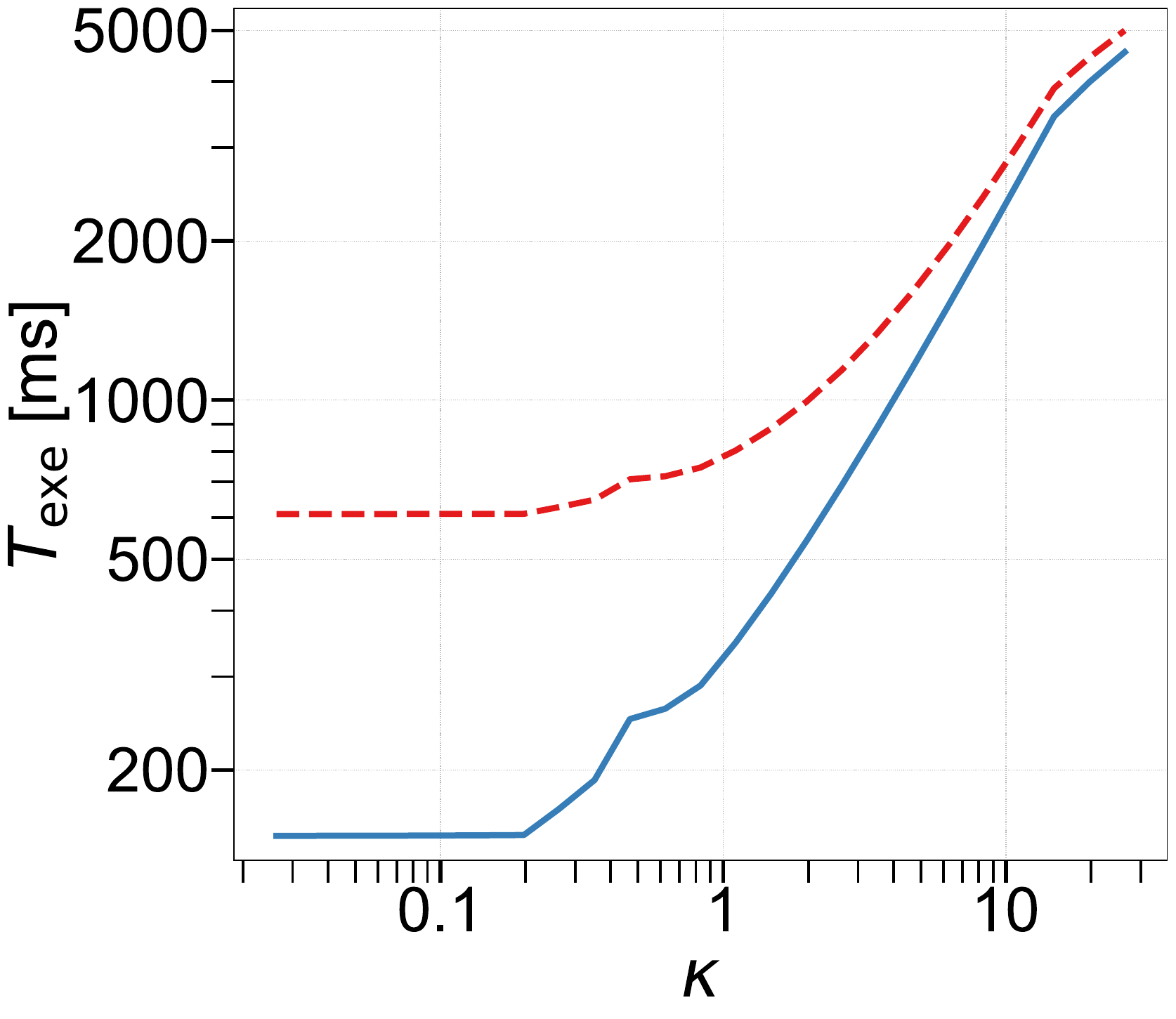}%
    \label{fig:two_anchor_sweep_kappa}%
  }
  \caption{
  Bottleneck switching in $T_{\mathrm{exe}}$ illustrated by one-dimensional sweeps of $R$, $d$, and $\kappa$ around the light/heavy reference configurations in Table~\ref{tab:even} ($S = 10$ shots). Plateaus and crossovers indicate transitions among $T_{\mathrm{ent}}$, $T_{\mathrm{cc}}$, and $T_{\mathrm{srv}}$.}
  \label{fig:two_anchor_sweeps}
\end{figure*}

We define two reference configurations using the endpoints of the baseline ranges in Table~\ref{tab:even}.
The light configuration is $(d_{\min}, R_{\max}, \kappa_{\min})$, and the heavy configuration is
$(d_{\max}, R_{\min}, \kappa_{\max})$.
From each reference configuration, we sweep one parameter at a time over a two-decade span
(from $0.1\times$ the light endpoint to $10\times$ the heavy endpoint) while fixing the other two. We report the execution time for a $10$-shot job.

The sweeps in Fig.~\ref{fig:two_anchor_sweeps} show bottleneck switching in $T_{\mathrm{exe}}$.
Increasing $R$ reduces $\Tent = K/R$ until another component limits $T_{\mathrm{exe}}$ (plateau)
(Fig.~\ref{fig:two_anchor_sweep_R}).
Increasing $d$ raises $\Tcc \propto d$, yielding a plateau-to-linear crossover once classical
communication dominates (Fig.~\ref{fig:two_anchor_sweep_d}).
Increasing $\kappa$ raises $\Tsrv$ and produces an analogous crossover when server execution
dominates (Fig.~\ref{fig:two_anchor_sweep_kappa}).

These crossovers show that the dominant latency component shifts across operating conditions, so ranking based on any single component is not consistent across regimes. We therefore evaluate multi-component telemetry-based ranking in the following experiments.

\subsection{Results} \label{sec:result}
\subsubsection*{Even regime}

We report mean oracle-normalized regret (Sec.~\ref{sec:setup}) over 1{,}500 instances for each $M$, with 95\% bootstrap confidence
intervals (CIs) over instances.
In the Even regime, $T_{\mathrm{srv}}$, $T_{\mathrm{cc}}$, and $T_{\mathrm{ent}}$ are of comparable magnitude, so no single component consistently dominates end-to-end runtime and the selection policy must trade off all three components. 
By construction, the Even-regime ranges in Table~\ref{tab:even} also straddle the classical--entanglement crossover $T_{\mathrm{cc}}=T_{\mathrm{ent}}$ (Fig.~\ref{fig:operating-map}), which induces bottleneck switching across candidates and makes ranking nontrivial.
\begin{figure}[htbp]
    \centering
    \includegraphics[width=1\linewidth]{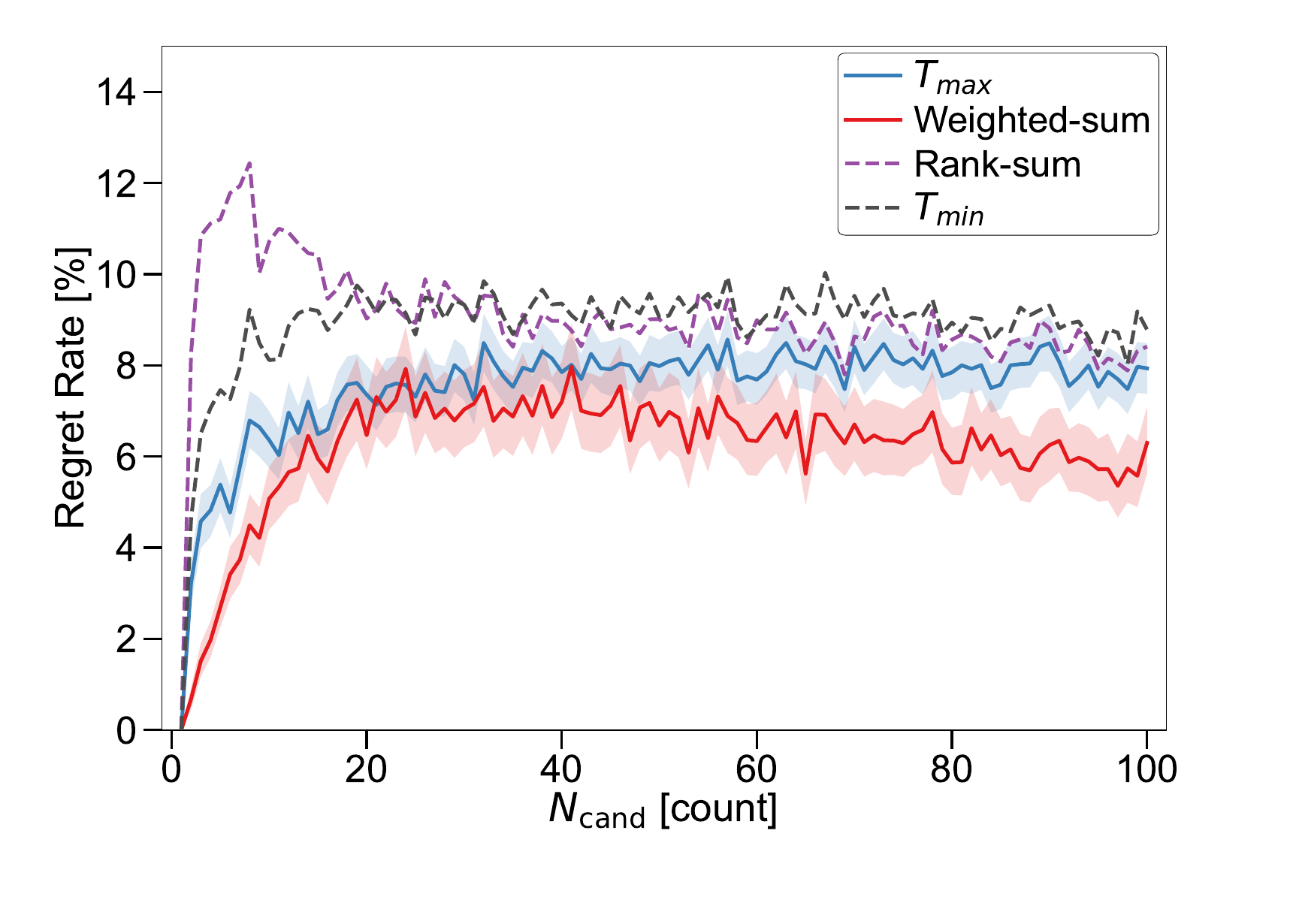}
    \caption{Mean oracle-normalized regret of server selection policies in the Even regime as a function of the number of candidate servers $M$.  Shaded bands indicate 95\% bootstrap confidence intervals.}
    \label{fig:even_bn}
\end{figure}
The Weighted-sum policy yields the lowest regret across $M$, with a mean regret of about 6\% (Fig.~\ref{fig:even_bn}). 
The $T_{\max}$ policy performs second best and maintains single-digit regret without weight calibration. 
Both $T_{\min}$ and Rank-sum yield higher regret. Overall, the results confirm that explicitly combining the latency components is essential for robust selection in the Even regime.

We additionally confirm that $T_{\max}$ is strongly rank-correlated with $T_{\mathrm{exe}}$ in the Even-regime pool (Spearman $\rho = 0.85$, Appendix~\ref{App:spear}).

\subsubsection*{Bottleneck-dominated regimes}
\begin{figure*}[t]
  \centering
  \subfloat[$T_{\mathrm{ent}}$-dominated]{%
    \includegraphics[width=0.33\linewidth]{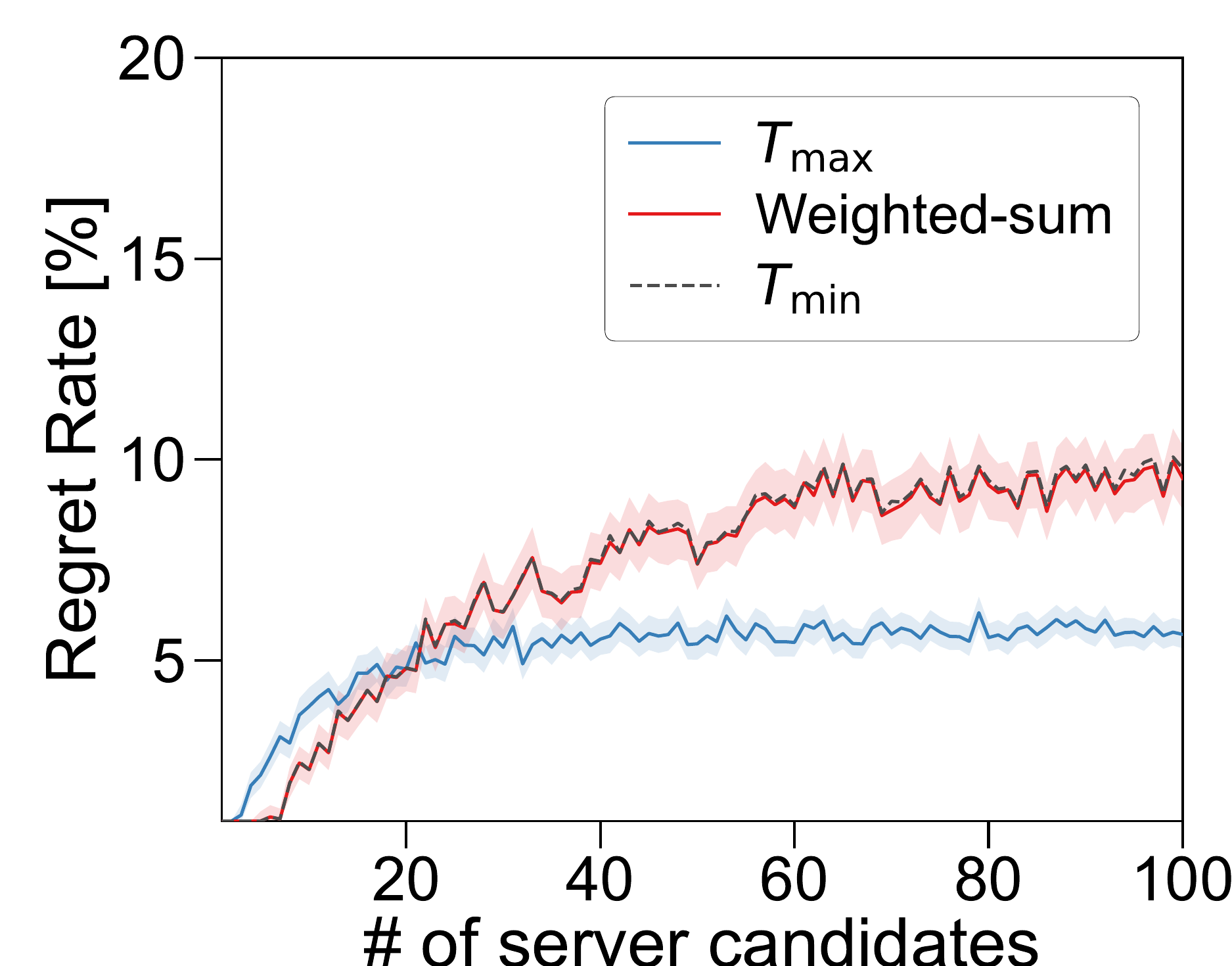}%
    \label{fig:bn_ent}%
  }\hfill
  \subfloat[$T_{\mathrm{cc}}$-dominated]{%
    \includegraphics[width=0.33\linewidth]{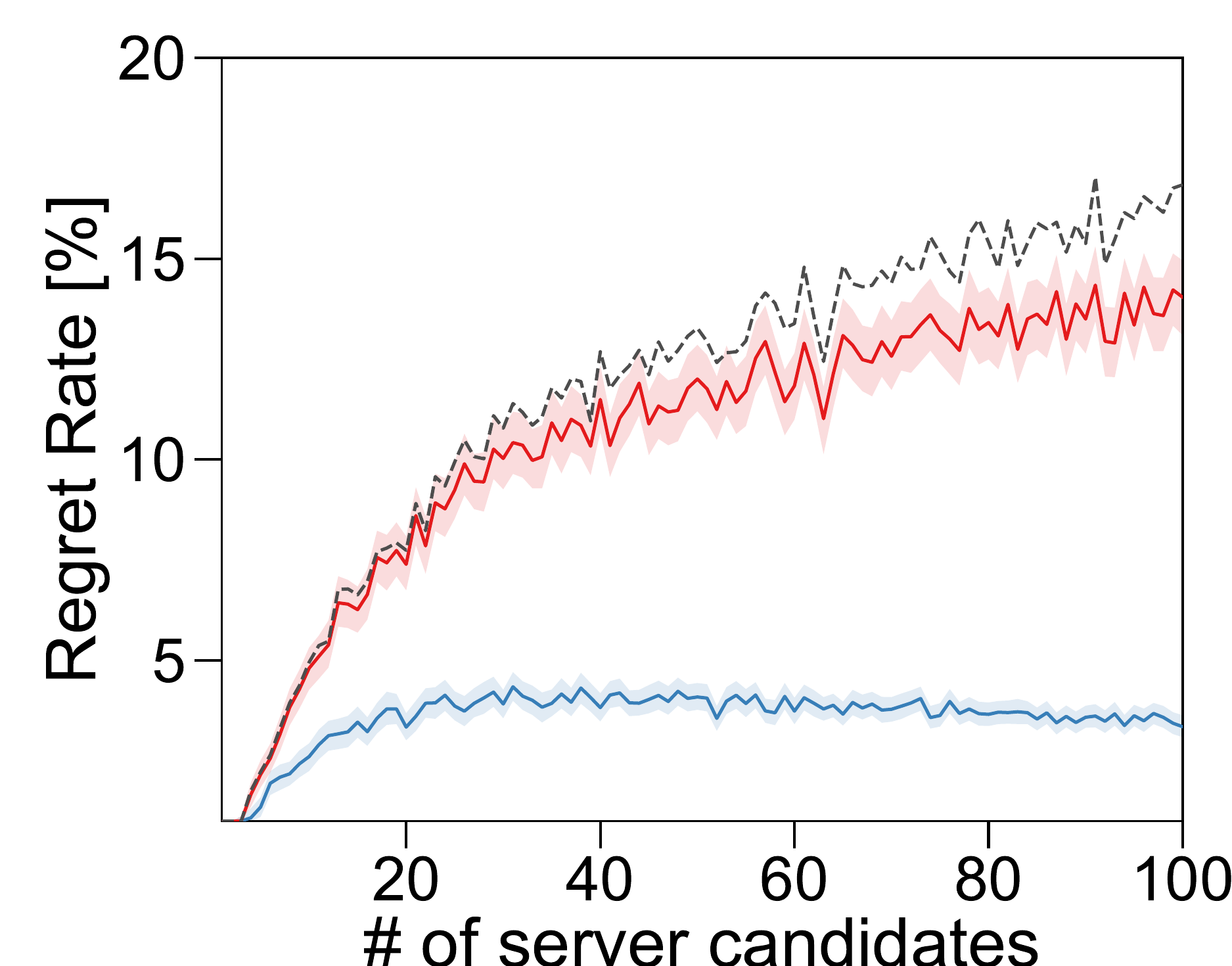}%
    \label{fig:bn_cc}%
  }\hfill
  \subfloat[$T_{\mathrm{srv}}$-dominated]{%
    \includegraphics[width=0.33\linewidth]{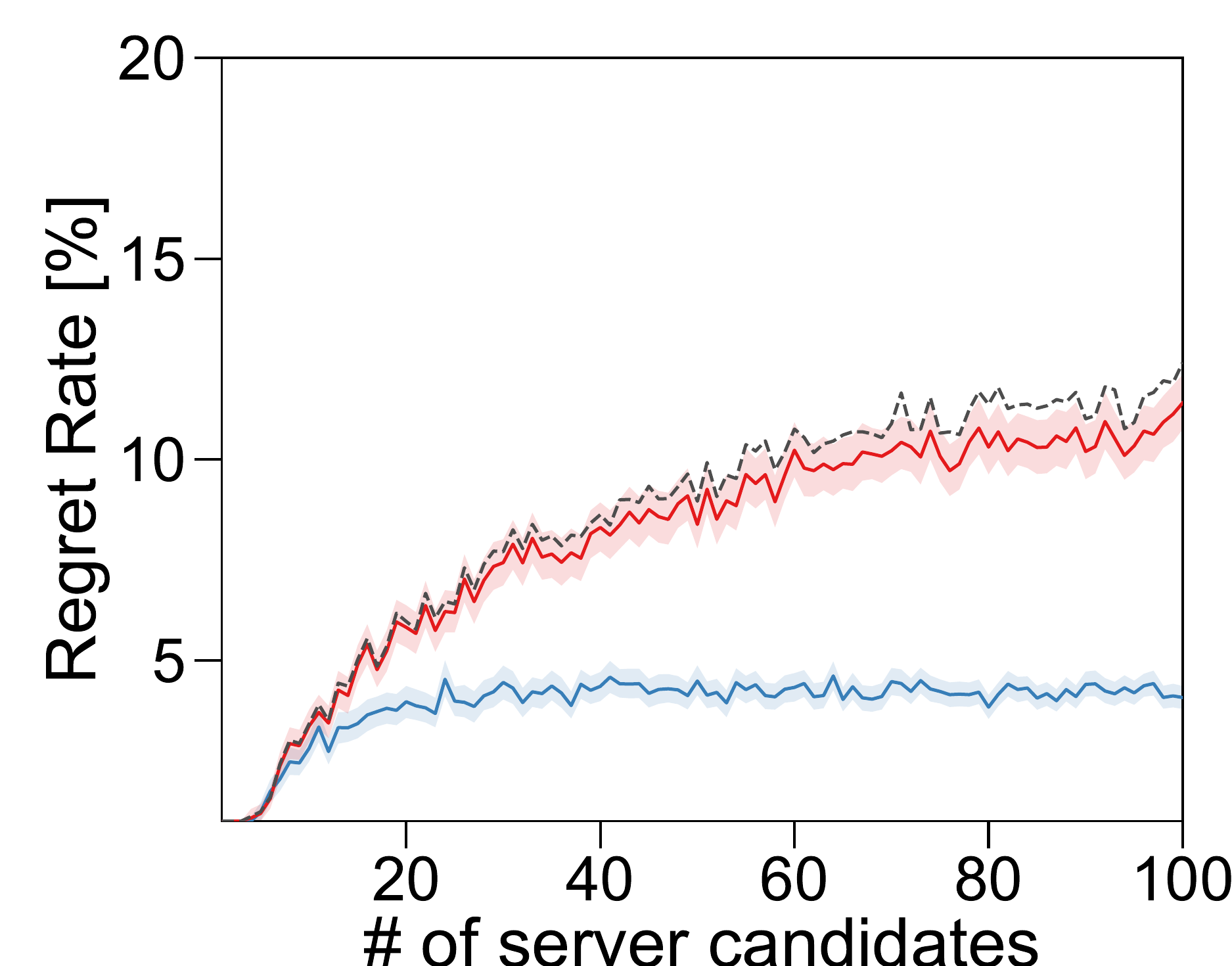}%
    \label{fig:bn_srv}%
  }
  \caption{
    Oracle-normalized regret of server-selection policies in bottleneck-dominated regimes versus the number of candidate servers $M$.
    Shaded bands indicate 95\% bootstrap confidence intervals.
    For visual clarity, we omit the Rank-sum baseline, whose regret is substantially larger and would compress the scale; the full curves including Rank-sum are reported in Appendix~\ref{app:ranksum_full} (Fig.~\ref{fig:bn_full}).
  }
  \label{fig:bn}
\end{figure*}

To probe bottleneck-dominated settings, we construct three regimes by scaling one parameter by a factor of 10 while holding the other two unchanged: we multiply $\kappa$ by 10 (server-computation-dominated), multiply $d$ by 10 (classical communication-dominated), and divide $R$ by 10 (entanglement-dominated). Regret across these regimes as a function of $M$ is summarized in Fig.~\ref{fig:bn}.

The $T_{\max}$ policy achieves the lowest mean regret (about 5\%) across the three bottleneck-dominated regimes. 
Weighted-sum and $T_{\min}$ yield comparable regret but are consistently higher than $T_{\max}$.

Notably, $T_{\max}$ outperforms the regime-trained Weighted-sum policy in all three bottleneck regimes, despite the latter behaving almost identically to $T_{\min}$. While a $10\times$ scaling makes one component dominant on average, the top candidates often exhibit near-ties on the dominant term, so regret is decided by the remaining delays. $T_{\max}$ remains sensitive to these secondary contributions, whereas a bottleneck-focused scalarization can suppress them and misorder near-optimal servers.
\subsubsection*{Robustness to temporal fluctuations} 

To evaluate robustness, we augment the propagation-limited one-way latency in optical fiber with an additive per-message jitter term. For each one-way classical message on the critical path, we model
\begin{equation}
t_{\mathrm{cc}}=\frac{d}{v_{\mathrm{fiber}}}+J,
\end{equation}
where $d$ is the channel length, $v_{\mathrm{fiber}}\approx2\times10^{5}\,\mathrm{km/s}$, and $J\ge 0$ is resampled for every message. We use an exponential jitter model
\begin{equation}
J \sim \mathrm{Exp}(\mu_J),
\end{equation}
with mean $\mu_J$ (i.e., $\mathrm{Exp}(\mu)$ denotes an exponential distribution with mean $\mu$).

We parameterize $\mu_J$ by the $0.999$-quantile $q_{0.999}$ of $J$:
\begin{equation}
q_{0.999}=\mu_J\ln(1000).
\end{equation}

Let $n_{\mathrm{cc}}$ denote the number of one-way classical messages on the critical path.
For an $S$-shot evaluation, the total jitter is 

\begin{equation}
J_{\mathrm{tot}}=\sum_{i=1}^{N}J_i \quad\text{for}\quad N=n_{\mathrm{cc}}S .
\end{equation}
Under the exponential model, 
the coefficient of variation of $J_{\mathrm{tot}}$ scales as $1/\sqrt{n_{\mathrm{cc}} S}$, 
so increasing $S$ reduces the relative impact of temporal variability on ranking under the same $q_{0.999}$.

We consider two representative settings that reflect different operational conditions: (i) Low (managed control-plane), $q_{0.999}=0.5\,\mathrm{ms}$, and (ii) High (cross-domain and contended), $q_{0.999}=20\,\mathrm{ms}$.

\begin{figure}[t]
  \centering
  \subfloat[High jitter ($q_{0.999}=20~\mathrm{ms}$)]{%
    \includegraphics[width=\linewidth]{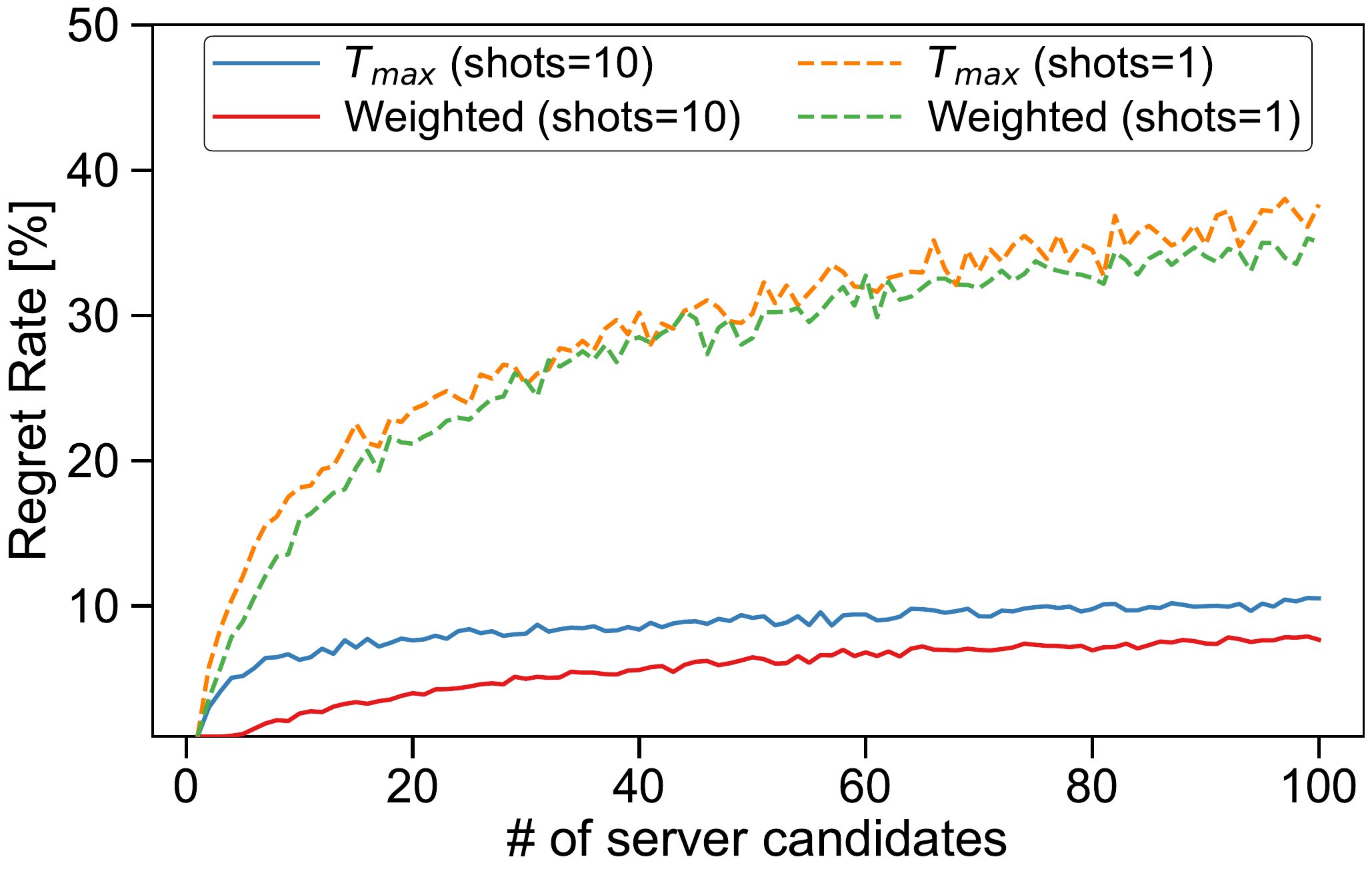}%
    \label{fig:jit_high}%
  }\par\medskip
  \subfloat[Low jitter ($q_{0.999}=0.5~\mathrm{ms}$)]{%
    \includegraphics[width=\linewidth]{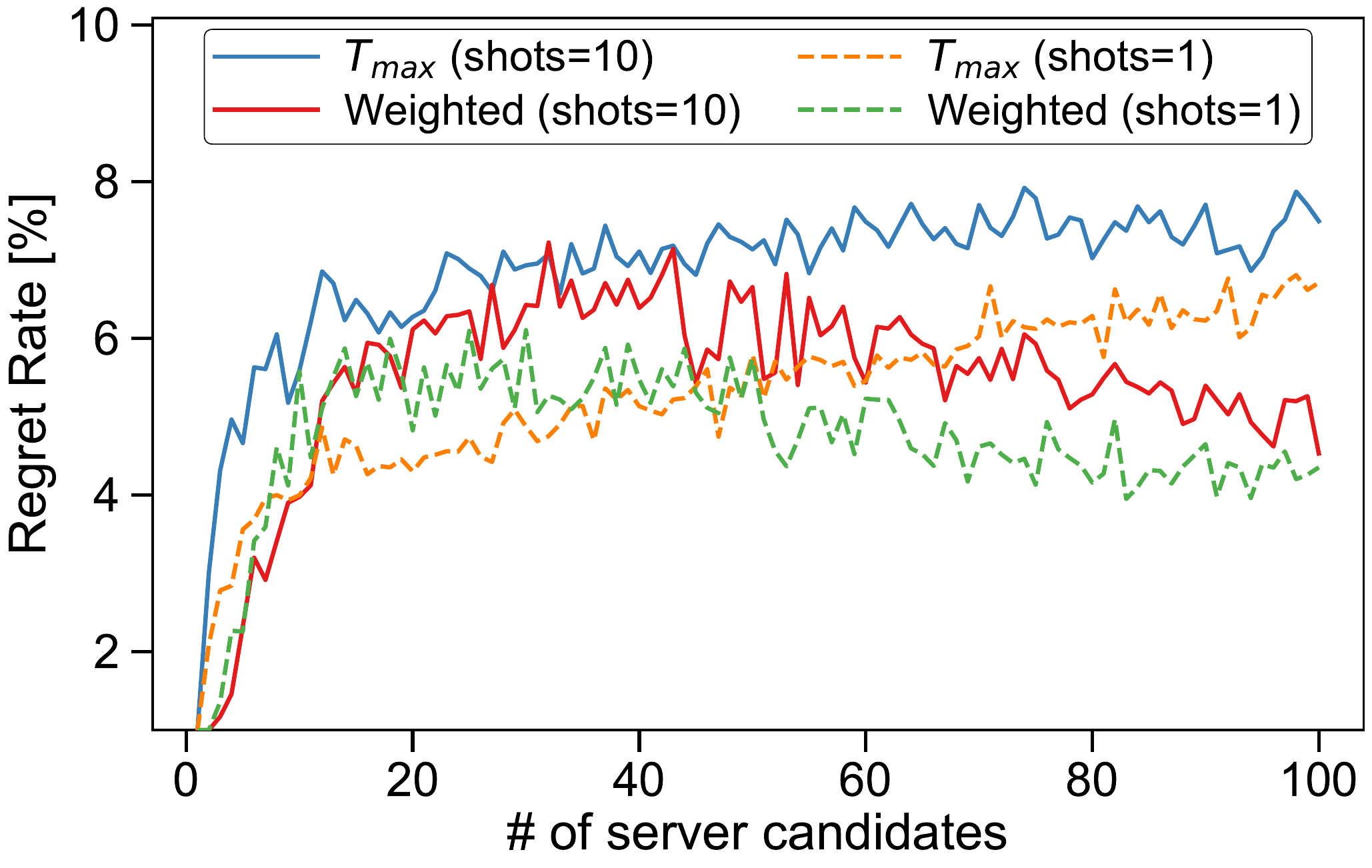}%
    \label{fig:jit_low}%
  }
  \caption{
    Oracle-normalized regret of $T_{\max}$ and Weighted-sum under classical communication latency jitter versus the number of candidate servers $M$.
    Solid/dashed curves correspond to $S=10$ and $S=1$.
  }
  \label{fig:jit}
\end{figure}

Temporal variability in classical communication latency degrades server selection performance, and the extent of the degradation depends on the shot count (Fig.~\ref{fig:jit}).

In the high-jitter setting (panel~(a), $q_{0.999}=20~\mathrm{ms}$), Weighted-sum generally attains lower regret than $T_{\max}$ for both $S=10$ (solid) and $S=1$ (dashed), while the gap remains modest and $T_{\max}$ remains close to the Weighted-sum curve.
However, under the same high-jitter condition, the single-shot case ($S=1$) exhibits substantial degradation as $M$ increases, reaching approximately $40\%$ at $M=100$ for both policies. In this failure regime, the two curves tend to converge, indicating reduced rankability from static telemetry under large message-level fluctuations. 
By contrast, in the low-jitter setting (panel~(b), $q_{0.999}=0.5~\mathrm{ms}$), regret remains in the single-digit range and is only weakly sensitive to $S$, suggesting that such managed control-plane jitter has limited impact on ranking. 
The pronounced $S$-dependence visible in panel~(a) is consistent with the coefficient-of-variation scaling $1/\sqrt{n_{\mathrm{cc}}S}$ of $J_{\mathrm{tot}}$ under the exponential jitter model. 

\subsubsection*{Weight Calibration for Telemetry-Based Ranking}
\label{sec:weight-vs-tmax}
Weighted-sum $T_w=\sum_{k\in\{\mathrm{srv,cc,ent}\}} w_kT_k$ generalizes $T_{\max}$ (uniform $w$). Here $w$ is derived from a Sobol sensitivity analysis from the assumed priors, computed offline and fixed (see Appendix~\ref{app:sobol} for details). In the Even regime, Weighted-sum attains the lowest regret ($\sim$6\%), while $T_{\max}$ remains single digit without tuning (Fig.~\ref{fig:even_bn}). In bottleneck-dominated regimes, where one component consistently drives the end-to-end latency, $T_{\max}$ achieves the lowest regret ($\sim$5\%; Fig.~\ref{fig:bn}). Under high classical-communication latency jitter, Weighted-sum is only marginally better, and in a single shot, both degrade strongly (Fig.~\ref{fig:jit}). Because the weights depend on the assumed priors, changes in operating conditions can reduce the benefit and may require recalibration. Thus, $T_{\max}$ is the robust default, whereas calibrated weighting is an optional refinement.

%% file: texs/eval_bottleneck.tex
\section{Bottleneck Identification} \label{sec:sobol}

\subsection{Requirement-Based Operating Map and Distance Envelopes}
\label{sec:operating-map}
The evaluation in Sec.~\ref{sec:setup} samples $(d, R, \kappa)$ independently to represent heterogeneous
candidate servers and network-service conditions. To connect these sweep ranges to deployment
scenarios and to generalize beyond a single benchmark instance, we derive a requirement-based
operating map in the $(d, R)$ plane using the protocol-level scalars $(n_{\mathrm{cc}}, K)$ defined in Sec.~\ref{sec:problem}.
In practical fiber links, the achievable end-to-end entanglement supply rate generally decreases with distance due to loss and is constrained in the repeaterless setting by fundamental rate--loss tradeoffs~\cite{Takeoka2014rateloss,Pirandola2017repeaterless}. Repeater design and network planning tools provide complementary guidance on feasible operating regions~\cite{Azuma2021tools,Azuma2023repeaters}.
Because the latency components scale linearly with the shot count $S$ when $(n_{\mathrm{cc}}, K)$ are
fixed per shot, the regime boundaries below are independent of $S$ and apply to both $S=1$ and
multi-shot jobs.

\subsubsection*{Classical--entanglement crossover boundary}
From Sec.~\ref{sec:problem} and the propagation-limited latency model in Sec.~\ref{sec:setup}, the feedforward delay and entanglement distribution time can be written directly as
\begin{equation}
T_{\mathrm{cc}} = n_{\mathrm{cc}} \frac{d}{v_{\mathrm{fiber}}}, \qquad
T_{\mathrm{ent}} = \frac{K}{R}.
\end{equation}

The crossover boundary $T_{\mathrm{cc}}=T_{\mathrm{ent}}$ yields the required entanglement distribution rate
\begin{equation}
R_{\mathrm{bal}}(d)=\frac{K\, v_{\mathrm{fiber}}}{n_{\mathrm{cc}}\, d}.
\label{eq:Rbal}
\end{equation}
Above this boundary ($R \gg R_{\mathrm{bal}}(d)$), entanglement distribution is fast enough that feedforward
latency tends to dominate. Below it ($R \ll R_{\mathrm{bal}}(d)$), entanglement distribution lies on the
critical path. Equivalently, for a network that can provide an end-to-end rate $R$, the distance at which
the classical and entanglement components become comparable is
\begin{equation}
d_{\mathrm{bal}}(R)=\frac{K\, v_{\mathrm{fiber}}}{n_{\mathrm{cc}}\, R}.
\label{eq:dbal}
\end{equation}

\subsection{PB-VQE instantiation and multiuser scaling}\label{sec:pbvqe-multiuser}

\subsubsection*{Operating map and PB-VQE instantiation}
\begin{figure}[t]
  \centering
  \includegraphics[width=1\linewidth]{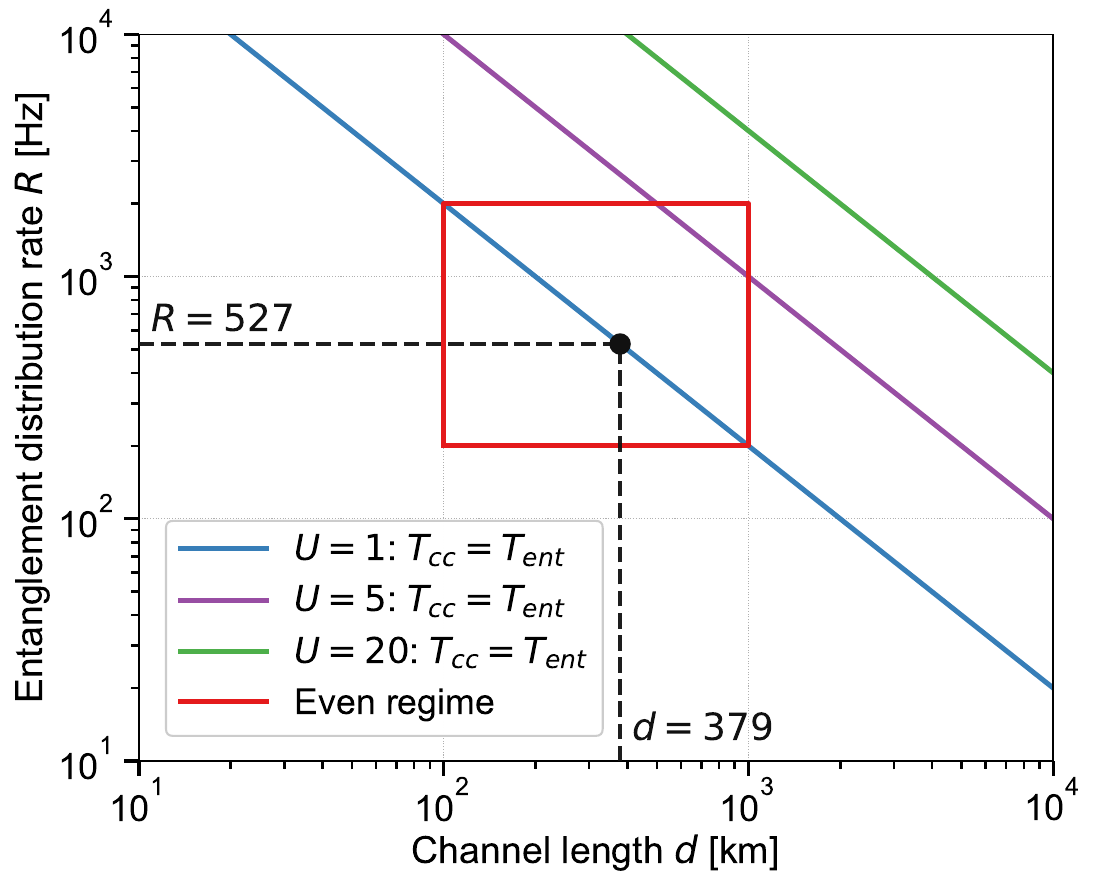}
  \caption{
  Requirement-based operating map in the $(d,R)$ plane for PB-VQE.
  The diagonal lines show the classical--entanglement crossover $T_{\mathrm{cc}}=T_{\mathrm{ent}}$ given by
  $R_{\mathrm{bal}}^{(U)}(d)=U\,R_{\mathrm{bal}}(d)=U\,v_{\mathrm{fiber}}/d$ (here $K=n_{\mathrm{cc}}=5$),
  for different multiuser sharing levels $U$ under $R_{\mathrm{eff}}=R/U$ ($U=1,5,20$).
  The red box indicates the Even-regime range (Table~\ref{tab:even}).
  The black dashed lines indicate the compute--network crossovers (Eqs.~\eqref{eq:kappa_cc}--\eqref{eq:kappa_ent}).
  The black dot marks the $\kappa=1$ balance point where $T_{\mathrm{srv}}\approx T_{\mathrm{cc}}\approx T_{\mathrm{ent}}$.
  }
  \label{fig:operating-map}
\end{figure}

Fig.~\ref{fig:operating-map} visualizes the resulting $(d,R)$ operating map for PB-VQE and highlights
how the evaluation ranges (Table~\ref{tab:even}) and the Sobol priors (Table~\ref{tab:sa-params}) cover distinct bottleneck regimes.

For our PB-VQE benchmark, $K=n_{\mathrm{cc}}=5$, hence $R_{\mathrm{bal}}(d)=v_{\mathrm{fiber}}/d$.
With $v_{\mathrm{fiber}}\approx 2\times 10^5\,\mathrm{km/s}$ and $d$ in km,
\begin{equation}
R_{\mathrm{bal}}(d)\approx \frac{2\times 10^5}{d}\;\; \mathrm{Hz}.
\label{eq:RbalPBVQE}
\end{equation}
Notably, the Even-regime ranges in Table~\ref{tab:even} ($d\in[100,1000]\,\mathrm{km}$ and $R\in[200,2000]\,\mathrm{Hz}$ ) are centered
around \eqref{eq:RbalPBVQE}, intentionally straddling the $T_{\mathrm{cc}}$--$T_{\mathrm{ent}}$ crossover to stress-test
selection under bottleneck switching.
In our benchmark, the telemetry-based approximation in Eq.~\eqref{eq:even} yields the per-shot components
\begin{align}
T_{\mathrm{srv}}\approx 9.48\,\kappa\;\mathrm{ms}, \quad
T_{\mathrm{cc}}\approx 0.025\, d\;\mathrm{ms}, \quad
T_{\mathrm{ent}}\approx \frac{5000}{R}\;\mathrm{ms},
\end{align}
with $d$ in km and $R$ in Hz. These yield simple crossover conditions:
\begin{align}
T_{\mathrm{srv}}=T_{\mathrm{cc}}\;\Rightarrow\; d \approx 379\,\kappa\;\mathrm{km}, \label{eq:kappa_cc}\\
T_{\mathrm{srv}}=T_{\mathrm{ent}}\;\Rightarrow\; R \approx \frac{527}{\kappa}\;\mathrm{Hz}.\label{eq:kappa_ent}
\end{align}

These relations provide a compact interpretation of when server-side acceleration (smaller $\kappa$)
versus network-layer provisioning (larger $R$ or smaller $d$) most effectively reduces end-to-end runtime.
These boundaries give quick deployment checks for PB-VQE:
(i) compare $R$ to $R_{\mathrm{bal}}(d)$ to determine whether $T_{\mathrm{ent}}$ or $T_{\mathrm{cc}}$ lies on the critical path, and
(ii) compute becomes limiting only when both network components are faster than the server, i.e., approximately $d \lesssim 379\kappa$ and $R \gtrsim 527/\kappa$.

\subsubsection*{Multiuser contention and load balancing}
In a multiuser (multi-client) setting, shared resources effectively rescale the parameters in the same operating map.
As a simple contention model, if $U$ clients share the entanglement distribution service along the relevant path,
we approximate the effective rate per client as $R_{\mathrm{eff}}=R/U$.
Under this model, the classical--entanglement crossover boundary scales linearly with concurrency,
\begin{equation}
R_{\mathrm{bal}}^{(U)}(d)=U\,R_{\mathrm{bal}}(d),
\end{equation}
which corresponds to a vertical shift of the boundary (cf. Fig.~\ref{fig:operating-map}).
Fig.~\ref{fig:operating-map} visualizes this shift: increasing $U$ moves the $T_{\mathrm{cc}}=T_{\mathrm{ent}}$ boundary upward by a factor of $U$,
so an operating point near the $\kappa=1$ balance point (black dot) becomes effectively more entanglement-limited unless the provisioned rate $R$ scales proportionally.
Similarly, contention for server-side compute can be captured by an effective slowdown (e.g., an increased $\kappa_{\mathrm{eff}}$ estimated from telemetry),
which expands compute-limited regions.
These scalings imply that load-aware telemetry (e.g., moving-average estimates of $R_{\mathrm{eff}}$ and $\kappa_{\mathrm{eff}}$) can be incorporated into the same $T_{\max}$ score with low overhead, enabling load-aware server ranking without run-time retuning.
This abstraction is consistent with the view that link-layer policies and scheduling determine the realized entanglement service rate under competing demands~\cite{Dahlberg2019linklayer, Pompili2022entanglementdelivery, Khatri2021policieselementary}.
Conversely, if entanglement resources are provisioned per client (so $R_{\mathrm{eff}}\approx R$), compute contention on a shared server can dominate. This can be captured as an increased $\kappa_{\mathrm{eff}}$ estimated from telemetry.

\subsection{Regime map of dominant bottlenecks and deployment implications}\label{sec:regime-map}
We provides an analytic operating map that compares the two network-side components ($T_{\mathrm{cc}}$ and $T_{\mathrm{ent}}$) in the $(d,R)$ plane in Fig.~\ref{fig:operating-map}.
However, end-to-end runtime depends jointly on $(\kappa,d,R)$ and on implementation-dependent overlap among server execution and network processes.

\begin{figure}[htbp]
    \centering    \includegraphics[width=1\linewidth]{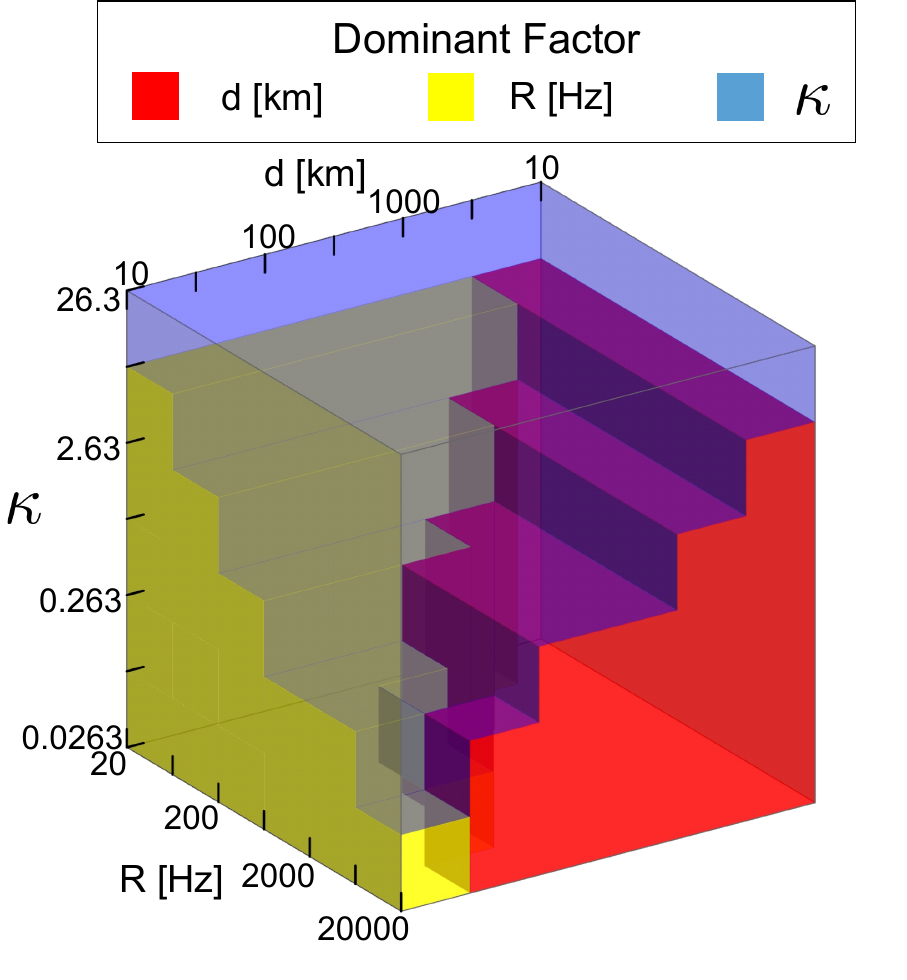}
    \caption{Dominant-factor map of Sobol total-effect sensitivity. Colors indicate the parameter with the largest total-effect index for runtime. Axes are $(\kappa, d, R)$ in log space over the ranges in Table~\ref{tab:sa-params}.}
    \label{fig:sobol}
\end{figure}

To visualize regime-dependent sensitivity beyond the analytic comparison, we estimate total-effect indices locally on the log-spaced grid used in Fig.~\ref{fig:sobol}, over parameter ranges in Table~\ref{tab:sa-params}.
Specifically, we partition the $(\kappa,d,R)$ space into log-spaced cells and, for each cell, draw Saltelli samples within the cell to estimate $(S_{T,\kappa}, S_{T,d}, S_{T,R})$.
Each color indicates the input with the largest total-effect index for runtime.

\begin{table}[htbp]
\centering
\caption{Sobol sensitivity-analysis inputs (independent log-uniform priors) and the log-space discretization in Fig.~\ref{fig:sobol}. We use expanded ranges derived from the Even-regime endpoints in Table~\ref{tab:even} by scaling the lower (upper) endpoint by $0.1$ ($10$).}
\label{tab:sa-params}
\small
\setlength{\tabcolsep}{4pt}
\renewcommand{\arraystretch}{1.05}
\begin{tabularx}{\columnwidth}{@{}
  >{\hsize=0.70\hsize\raggedright\arraybackslash}X
  >{\hsize=0.30\hsize\raggedright\arraybackslash}X
@{}}
\hline
\textbf{Parameter} & \textbf{Range} \\
\hline
Operation-duration scaling factor $\kappa$ & $0.0263$--$26.3$ \\
Entanglement distribution rate $R$ & $20$--$20{,}000\,\mathrm{Hz}$ \\
Channel length $d$ & $10$--$10{,}000\,\mathrm{km}$ \\
\hline
\end{tabularx}
\end{table}

The Sobol total-effect indices for runtime are summarized in Fig.~\ref{fig:sobol}. Colors indicate the parameter with the largest total-effect index. The dominant constraint shifts across operating regimes. At short channel lengths, the runtime is primarily limited by the entanglement distribution rate $R$. At long channel lengths, classical communication latency $\Tcc$ over fiber dominates. $\Tcc$ scales with channel length $d$ and the number of one-way classical communication messages $\ncc$ on the critical path, reaching tens to hundreds of $\mathrm{ms}$ for intercontinental links.

In entanglement-limited regimes, the total-effect index for $R$ dominates, as expected from $T_{\mathrm{ent}}=K/R$ on the critical path.
In this region, improving the entanglement distribution service (larger $R$ at the required fidelity) yields the largest runtime reductions.
In long-haul regimes, the total-effect index for $d$ dominates, consistent with feedforward latency $T_{\mathrm{cc}}=n_{\mathrm{cc}}d/v_{\mathrm{fiber}}$ becoming limiting.
In intermediate regimes where $T_{\mathrm{srv}}$ becomes comparable, improving server-side execution (smaller $\kappa$) and network-layer provisioning
(larger $R$ or smaller $d$) become substitutable levers.

These regime-dependent priorities also help interpret why the additive score $T_{\max}$ yields robust selection performance in Sec.~\ref{sec:evaluation}:
the sum is monotone in each latency component and is dominated by the bottleneck term in bottleneck-dominated regimes,
while still providing a simple, calibration-free scalarization of cross-layer trade-offs near crossover boundaries.

%% file: texs/conclusion.tex
\section{Conclusion}\label{sec:conclusion}
We address application-layer server selection for Quantum Internet applications whose end-to-end runtime couples server execution, feedforward classical latency, and entanglement supply, with protocol-dependent overlap and regime-dependent bottleneck switching.
We proposed the lightweight telemetry-based score $T_{\max}$ for conservative online ranking and, using large-scale NetSquid discrete-event simulations of an implementation-oriented PB-VQE benchmark, showed that $T_{\max}$ achieves consistently low oracle-normalized regret across crossover and bottleneck-dominated regimes, including temporal jitter scenarios.

Beyond selection accuracy, Sec.~\ref{sec:operating-map} derives a requirement-based operating map in the $(d,R)$ plane and simple distance envelopes that relate network requirements to protocol-level counts, while Sec.~\ref{sec:pbvqe-multiuser} translates these regimes into provisioning implications: which telemetry matters most in each region and how multi-user contention shifts the classical--entanglement boundary via $R_{\mathrm{eff}}=R/U$.
Finally, Sec.~\ref{sec:regime-map} complements the analytic map with a regime map of dominant bottlenecks (Fig.~\ref{fig:sobol}) via local Sobol total-effect indices, indicating which telemetry dimension ($d$, $R$, or $\kappa$) is the most effective lever in each operating region.
Together, these results provide both a practical selection rule and interpretable deployment guidance for emerging Quantum Internet services.

\subsection*{Limitations and outlook}
To keep client-side decision overhead low, we assume access to coarse-grained telemetry on
server-side computational performance, classical communication latency, and entanglement-distribution rate.
Our simulations focus on a single active client and omit queuing delays and multi-client contention.
We also treat the application-level accuracy constraint as a precondition on the candidate set.
An important direction for future work is to develop probabilistic and load-aware predictors that
capture temporal correlations, scheduling, and contention in multi-client deployments.

%% file: texs/appendix.tex
\appendices
\section{Implementation-Oriented PB-VQE} \label{sec:pbvqe}
This appendix specifies the modified PB-VQE benchmark used in Sec.~\ref{sec:evaluation}, including the protocol-dependent counts $(n_{\mathrm{cc}},K)$ on the critical path.
\subsection{Protocol Design}

\begin{figure}[htbp]
    \centering
    \includegraphics[width=1\linewidth]{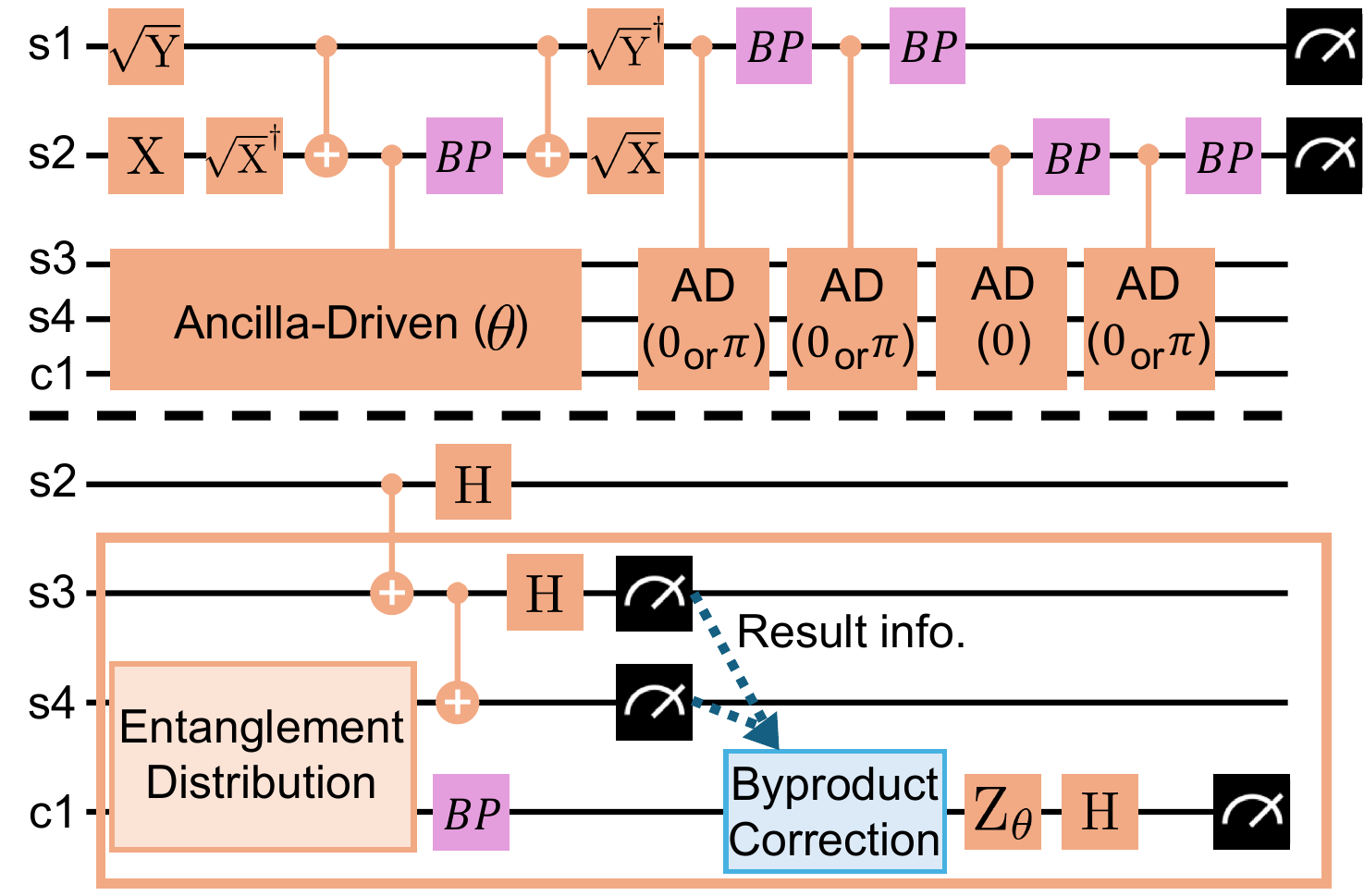}
    \caption{Modified parameter-blind VQE circuit (top) and ancilla-driven gate decomposition (bottom). Server qubits are labeled s1--s4, and c1 is a client ancilla for teleportation and byproduct (BP) correction~\cite{anders2010ancilla}. Measurement labels $\theta$ and 0/$\pi$ denote the angles chosen at each step.}
    \label{fig:pbvqe-circuit}
\end{figure}

The parameter-blind VQE (PB-VQE) protocol is a client--server variant of VQE, in which the server always executes a fixed, parameter-independent circuit. In contrast, the client updates the variational parameters based on measurement outcomes~\cite{BlindparametaVQE}. This ensures that the variational parameters remain hidden from the server, thereby achieving parameter blindness with significantly reduced communication compared to conventional blind quantum computation (BQC) protocols such as BFK~\cite{broadbent2009universalBFK} or MF~\cite{morimae2013blind}.
We propose a modified PB-VQE protocol, tailored for implementation over the Quantum Internet. We integrate quantum teleportation to improve robustness against channel loss. We add four ancilla-driven computations to correct teleportation byproducts and preserve parameter blindness (Fig.~\ref{fig:pbvqe-circuit}). This modification increases the feasibility of the protocol in realistic, lossy network conditions. 
Ancilla-driven (AD) decomposition consumes one Bell pair and incurs one one-way feedforward classical message (the Bell-measurement outcomes) per block, we have $n_{\mathrm{cc}}=K=5$ per shot. Tracing the server-side critical path of the fixed circuit instance in Fig.~\ref{fig:pbvqe-circuit} yields $T_{\mathrm{srv}}\approx 9.48~\mathrm{ms}$ per shot.

\subsection{Measurement Update Policy}
\label{sec_appMUP}
In the modified PB-VQE protocol (Fig.~\ref{fig:pbvqe-circuit}), the server executes 
a fixed circuit 
AD($\theta$) $\to$ AD(0/$\pi$) $\to$ AD(0/$\pi$) $\to$ AD(0) $\to$ AD(0/$\pi$), 
while the client updates the measurement angles and 
measures the ancilla after each block (labeled 1–5). 
At 0/$\pi$ stages, the angle is set to either 0 or $\pi$. 
Each ancilla-driven step applies, conditioned on the ancilla 
outcome $b$, the operation $(X^b) H R_z(\alpha)$ to the register.

Quantum teleportation produces two Bell-measurement bits $(z,x)$, 
which induce the Pauli byproduct $Z^z X^x$. 
The measurement angle is updated in the Pauli frame as
\begin{equation}
  \alpha \;\mapsto\; (-1)^x \alpha + z\pi \pmod{2\pi}.
  \label{eq:pauli-update}
\end{equation}
This corresponds to the byproduct correction in Fig.~\ref{fig:pbvqe-circuit}. 

For the five ancilla-driven (AD) blocks, 
the measurement angle $\theta_t$ is set according to the gate type:
\begin{equation}
\theta_t =
\begin{cases}
  \theta   & \text{for AD}(\theta), \\[6pt]
  \pi m    & \text{for AD}(0/\pi), \\[6pt]
  0        & \text{for AD}(0),
\end{cases}
\end{equation}
where $m \in \{0,1\}$ is the measurement outcome from the previous step.

This explicit update rule ensures that byproducts are corrected 
consistently at every stage. 
With these corrections in place, we analyze 
the blindness properties of the modified PB-VQE protocol.

\subsection{Blindness Properties}
\subsubsection*{Parameter blindness}
A single ancilla-driven circuit produces one classical bit  
\(m\in\{0,1\}\).  Conditional on \(m\), the register undergoes one of the two gates  
\begin{align}
U_0 = H\,R_z(\theta),\qquad
U_1 = X\,H\,R_z(\theta).
\end{align}
For the Fig.~\ref{fig:pbvqe-circuit} ancilla-driven decomposition (ancilla initialized in $|0\rangle$), the register Kraus operators are
\begin{align}
  K_0 &= \tfrac{1}{\sqrt{2}}\, H R_z(\theta), \nonumber\\
  K_1 &= \tfrac{1}{\sqrt{2}}\, H Z R_z(\theta)
      = \tfrac{1}{\sqrt{2}}\, X H R_z(\theta),
\end{align}

so $p(m)=\Tr(K_m\rho K_m^\dagger)=\tfrac12$ and the server must average uniformly over the hidden $m$.
Let \(\rho\) be the register state immediately before the ancilla-driven circuit step.  
From the server’s point of view (the measurement outcome \(m\) is hidden), the
post-measurement state is
\begin{align}
\rho_{\mathrm{srv}}
  &= \tfrac12\bigl(U_0\rho U_0^\dagger + U_1\rho U_1^\dagger\bigr)\nonumber\\
  &= \tfrac12
    H R_z(\theta)\bigl(\rho + Z\rho Z\bigr)R_z^\dagger(\theta)H^\dagger.
\end{align}

The prefactor $\tfrac12(\rho+Z\rho Z)$ commutes with $Z$.
Since $R_z(\theta)=e^{-i\theta Z/2}$ is generated by $Z$, it follows that
\begin{align}
  R_z(\theta)\bigl(\rho+Z\rho Z\bigr)R_z^\dagger(\theta)
    &= \rho+Z\rho Z, \\
  \rho_{\mathrm{srv}}
    &= \tfrac12\, H\bigl(\rho+Z\rho Z\bigr) H^\dagger .
\end{align}
is independent of~$\theta$, and no POVM applied by the server can reveal any
information about the variational parameter. We thus obtain parameter blindness.

\subsubsection*{Output blindness}

Immediately before the final computational-basis measurement, we apply two consecutive ancilla-driven steps. Let $\mathcal{E}_\theta$ denote the effective single-step channel in the server's reduced description, obtained by averaging over the unknown Pauli byproduct $Z^b$ with $b\in\{0,1\}$:
\begin{align}
  \mathcal{E}_\theta(\rho)
  = \tfrac12\,H R_z(\theta)\bigl(\rho + Z\rho Z\bigr)R_z^\dagger(\theta)H^\dagger .
\end{align}
For $\rho$ of the form $\tfrac12(I+r_x X+r_y Y+r_z Z)$, we have
\begin{align}
  \mathcal{E}_\theta(\rho)
  &= \tfrac{1}{2}\,H R_z(\theta)\bigl(I + r_z Z\bigr)R_z^\dagger(\theta)H^\dagger \nonumber\\
  &= \tfrac12\bigl(I + r_z X\bigr),
\end{align}
which is independent of \(\theta\). Applying the same channel once more gives
\begin{align}
  \mathcal{E}_\theta\!\left(\mathcal{E}_\theta(\rho)\right)
  &= \tfrac12\,H R_z(\theta)\Bigl(\mathcal{E}_\theta(\rho) + Z\,\mathcal{E}_\theta(\rho)\,Z\Bigr)R_z^\dagger(\theta)H^\dagger \nonumber\\
  &= \tfrac{I}{2}.
\end{align} 
Thus, two hidden steps implement the completely depolarizing channel on each qubit from the server's perspective. Applying this to all \(n\) qubits yields
\begin{align}
  \rho_{\mathrm{final}} = \frac{I}{2^{n}}
\end{align}
for any input \(\rho_{\mathrm{in}}\). Therefore, the server cannot distinguish the logical output or any intermediate branch of the computation, and output blindness holds.

\section{Validity of telemetry-based predictors} \label{App:spear}

\begin{figure}[htbp]
    \centering
    \includegraphics[width=1\linewidth]{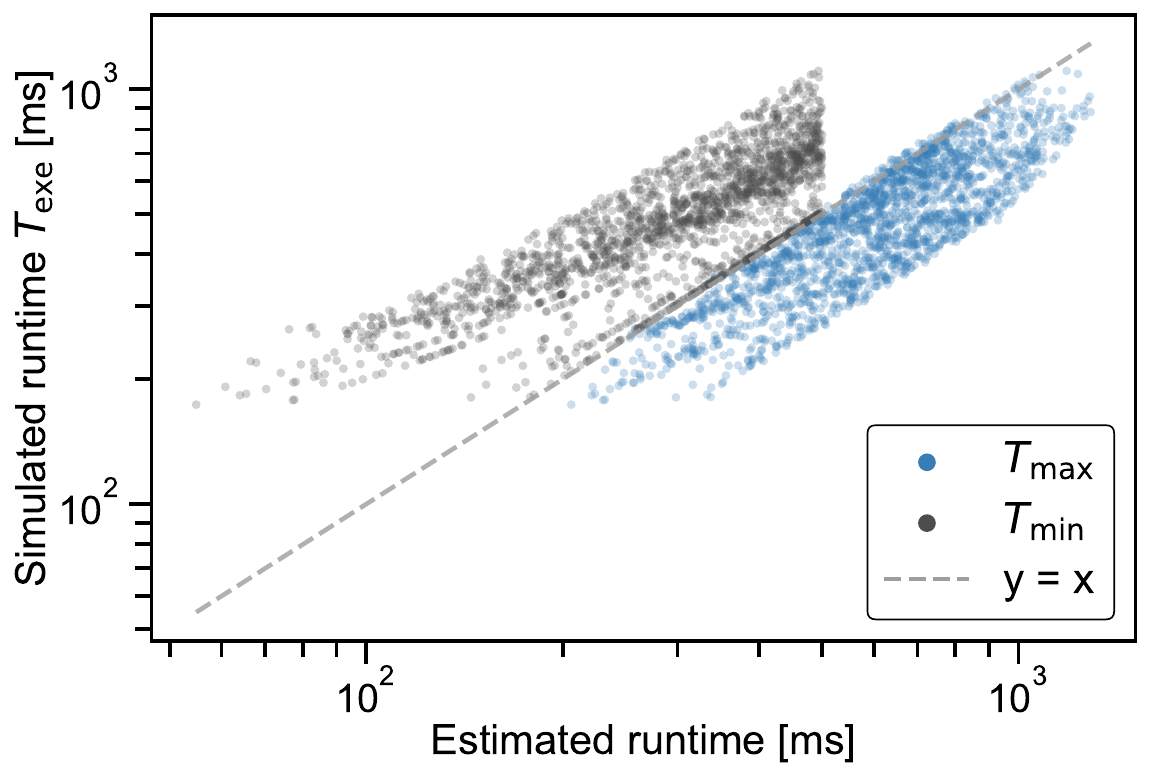}
    \caption{Spearman's rank correlation between telemetry-based runtime estimates and the simulated execution time $T_{\mathrm{exe}}$ over $2000$ server configurations sampled from the even-regime pool (Table~\ref{tab:even}). Both axes are on logarithmic scales. The rank correlation is $\rho=0.85$ for $T_{\max}$ and $\rho=0.79$ for $T_{\min}$.}
    \label{fig:rank_sanity}
\end{figure}

We compute Spearman's rank correlation coefficient between each telemetry-based estimate and the simulated execution time $T_{\mathrm{exe}}$ over the even-regime pool. 
Spearman’s rank correlation is $\rho=0.85$ for $T_{\max}$ and $\rho=0.79$ for $T_{\min}$ (Fig.~\ref{fig:rank_sanity}).
The higher coefficient for $T_{\max}$ suggests a stronger monotonic relationship with $T_{\mathrm{exe}}$ in this setting. Selection performance is evaluated directly in Sec.~\ref{sec:evaluation}.

\section{Full bottleneck-dominated regret curves including Rank-sum}
\label{app:ranksum_full}

\begin{figure*}[t]
  \centering
  \subfloat[$T_{\mathrm{ent}}$-dominated]{%
    \includegraphics[width=0.33\linewidth]{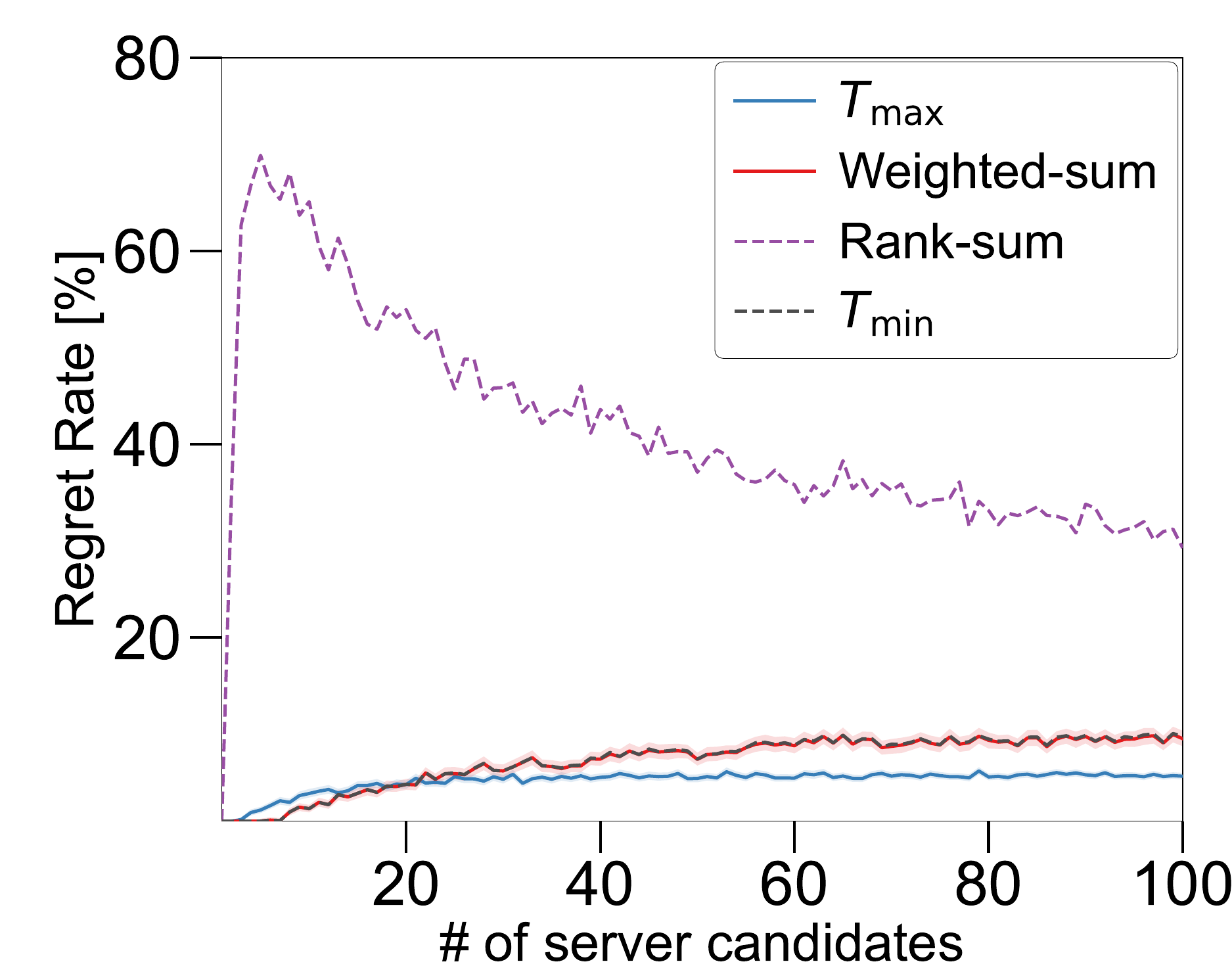}%
  }\hfill
  \subfloat[$T_{\mathrm{cc}}$-dominated]{%
    \includegraphics[width=0.33\linewidth]{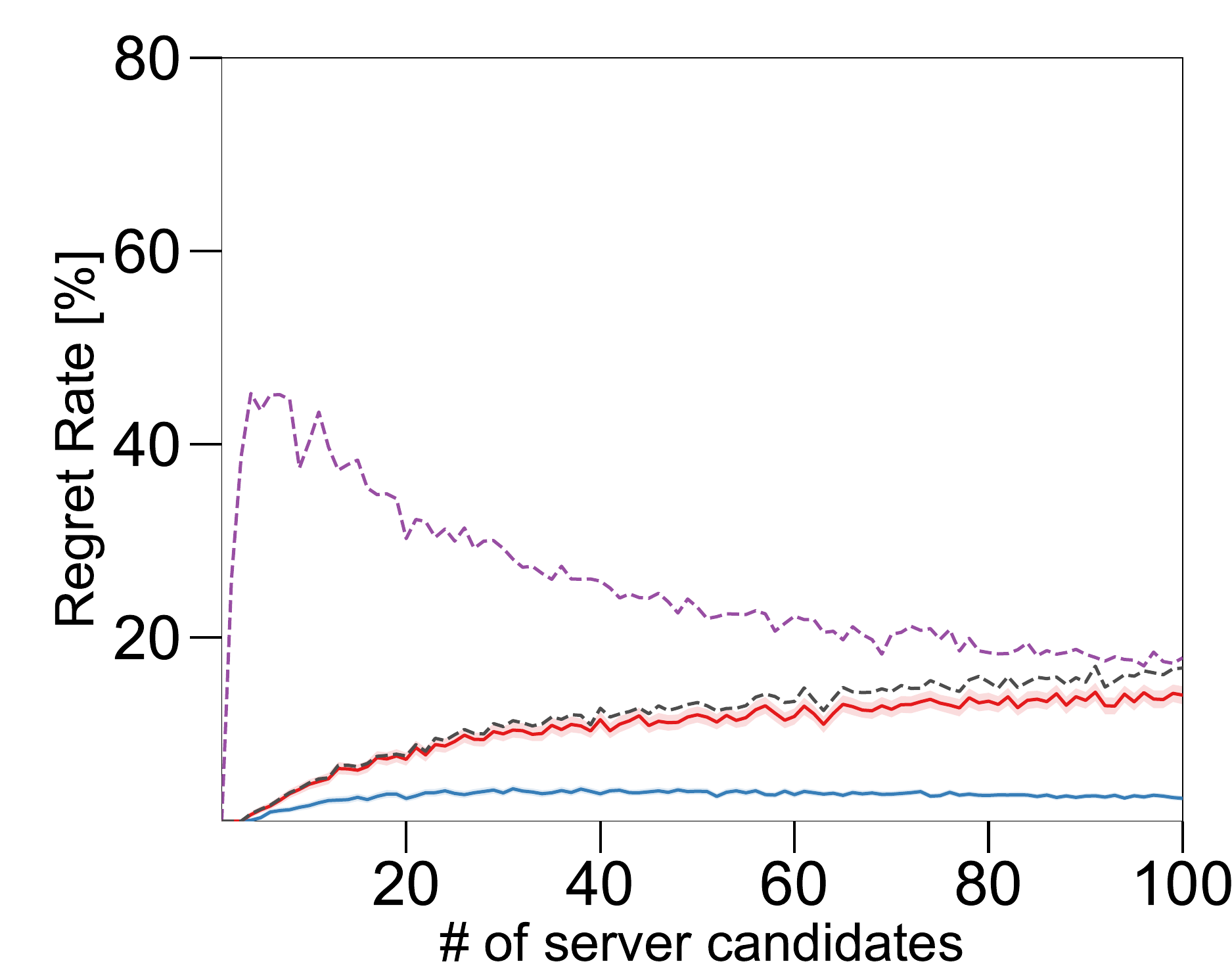}%
  }\hfill
  \subfloat[$T_{\mathrm{srv}}$-dominated]{%
    \includegraphics[width=0.33\linewidth]{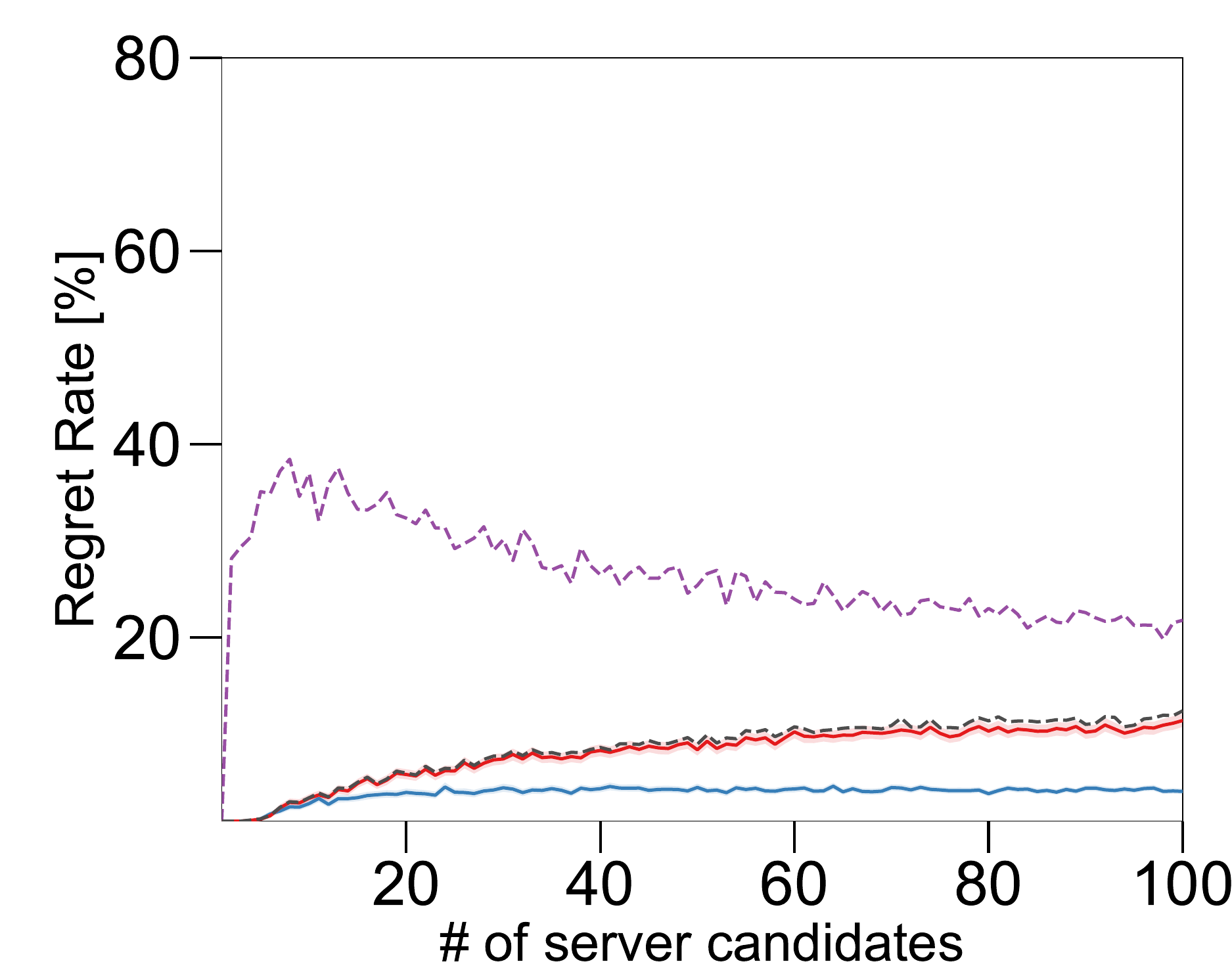}%
  }
  \caption{
    Oracle-normalized regret of server-selection policies in bottleneck-dominated regimes versus the number of candidate servers $M$.
    Shaded bands indicate 95\% bootstrap confidence intervals.
  }
  \label{fig:bn_full}
\end{figure*}

In Fig.~\ref{fig:bn}, we omit the Rank-sum baseline for readability because its regret is substantially larger than the other policies in bottleneck-dominated regimes, which would otherwise compress the vertical scale and obscure the differences among $T_{\max}$, $T_{\min}$, and Weighted-sum. Fig.~\ref{fig:bn_full} reports the full curves including Rank-sum for completeness.

\section{Sobol Total-Effect Indices}\label{app:sobol}
We identify bottlenecks using variance-based global sensitivity analysis with Sobol indices~\cite{sobol1990sensitivity}. We use Sobol analysis as a screening tool and report total-effect indices, which provide a conservative ranking of influential inputs when the inputs are treated as independent.

Let $Y$ be the output metric and $\mathbf{X}=(X_1,\ldots,X_p)$ the inputs.
Denote by $\mathbf{X}_{\sim i}=(X_1,\ldots,X_{i-1},X_{i+1},\ldots,X_p)$ the set of all inputs except $X_i$.
The total-effect index for $X_i$ is
\begin{equation}
S_{T_i} \;=\; 1 \;-\; \frac{\mathrm{Var}_{\mathbf{X}_{\sim i}}\!\left(\mathbb{E}\!\left[Y \mid \mathbf{X}_{\sim i}\right]\right)}{\mathrm{Var}(Y)}.
\end{equation}

We draw samples using the Saltelli scheme~\cite{saltelli2010variance} with quasi-random sampling over the log-uniform ranges summarized in Table~\ref{tab:sa-params}. We use $N=128$ base samples in the Saltelli design. Each sample fixes the server-side operation-duration scaling factor $\kappa$,
 channel length $d$, and entanglement distribution rate $R$.
 For each configuration, we run PB-VQE and record runtime, then estimate $S_{T_i}$. 

We set the output as the $S$-shot end-to-end runtime $Y = T_{\mathrm{exe}}$ and choose the Sobol inputs as $X = (\kappa, d, R)$, which correspond directly to the three latency components $(T_{\mathrm{srv}}, T_{\mathrm{cc}}, T_{\mathrm{ent}})$ through the runtime decomposition in Sec.~\ref{sec:setup}. 
We interpret the corresponding total-effect indices as $S_{T,\mathrm{srv}}$, $S_{T,\mathrm{cc}}$, and $S_{T,\mathrm{ent}}$, and use them to form an offline-calibrated weight vector for the Weighted-sum baseline in Sec.~\ref{sec:policy}:
\begin{align}
(w_{\mathrm{srv}},\, w_{\mathrm{cc}},\, w_{\mathrm{ent}})
 = \frac{(S_{T,\mathrm{srv}},\, S_{T,\mathrm{cc}},\, S_{T,\mathrm{ent}})}
 {S_{T,\mathrm{srv}} + S_{T,\mathrm{cc}} + S_{T,\mathrm{ent}}}.
\end{align}
These weights are fixed across all instances to keep the online decision-time overhead minimal. 